\documentclass[10pt,twocolumn]{IEEEtran}

\usepackage{amsfonts,amssymb,amsmath}

\newtheorem{Def}{Definition}
\newtheorem{Lem}{Lemma}
\newtheorem{Thrm}{Theorem}
\newtheorem{Cor}{Corollary}

\begin{document}
\title{Unconditionally Secure Quantum Key Distribution In Higher Dimensions}

\author{H. F. Chau, Member, IEEE\thanks{H. F. Chau is with the Department of
 Physics, University of Hong Kong, Pokfulam Road, Hong Kong. (E-mail:
 hfchau@hkusua.hku.hk)}}
\maketitle

\begin{abstract}
 In search of a quantum key distribution scheme that could stand up for more
 drastic eavesdropping attack, I discover a prepare-and-measure scheme using
 $N$-dimensional quantum particles as information carriers where $N$ is a prime
 power. Using the Shor-Preskill-type argument, I prove that this scheme is
 unconditional secure against all attacks allowed by the laws of quantum
 physics. Incidentally, for $N = 2^n > 2$, each information carrier can be
 replaced by $n$ entangled qubits. And in this case, I discover an
 eavesdropping attack on which no unentangled-qubit-based prepare-and-measure
 quantum key distribution scheme known to date can generate a provably secure
 key. In contrast, this entangled-qubit-based scheme produces a provably secure
 key under the same eavesdropping attack whenever $N \geq 16$. This
 demonstrates the advantage of using entangled particles as information
 carriers to combat certain eavesdropping strategies.
\end{abstract}


\section{Introduction \label{Sec:Intro}}
\PARstart{K}{ey} distribution is the art of sharing a secret key between two
 cooperative players Alice and Bob in the presence of an eavesdropper Eve. If
 Alice and Bob distribute their key by exchanging classical messages only, Eve
 may at least in principle wiretap their conversations without being caught.
 So, given unlimited computational resources, Eve can crack the secret key. In
 contrast, in any attempt to distinguish between two non-orthogonal states,
 information gain is only possible at the expenses of disturbing the state
 \cite{NC}. Therefore, if Alice and Bob distribute their secret key by sending
 non-orthogonal quantum signals, any eavesdropping attempt will almost surely
 affect their signal fidelity. Consequently, a carefully designed quantum key
 distribution (QKD) scheme allows Alice and Bob to accurately determine the
 quantum channel error rate, which in turn reflects the eavesdropping rate. If
 the estimated quantum channel error rate is too high, Alice and Bob abort the
 scheme and start all over again. Otherwise, they perform certain privacy
 amplification procedures to distill out an almost perfectly secure key
 \cite{biasedbb84,lochauqkdsec,mayersjacm,gottloreview,gisinreview}. Therefore,
 it is conceivable that a provably secure QKD scheme exists even when Eve has
 unlimited computational power.

 With this belief in mind, researchers proposed many QKD schemes
 \cite{gisinreview}. These schemes differ in many ways such as the Hilbert
 space dimension of the quantum particles used, as well as the states and bases
 Alice and Bob prepared and measured. The first QKD scheme, commonly known as
 BB84, was invented by Bennett and Brassard \cite{bb84}. In BB84, Alice
 randomly and independently prepares each qubit in one of the following four
 states: $|0\rangle$, $|1\rangle$ and $(|0\rangle \pm |1\rangle)/\sqrt{2}$, and
 sends them to Bob. Upon reception, Bob randomly and independently measures
 each qubit in either the $\{ |0\rangle, |1\rangle \}$ or $\{ (|0\rangle \pm |1
 \rangle)/\sqrt{2} \}$ bases \cite{bb84}. In short, BB84 is an experimentally
 feasible prepare-and-measure scheme involving the transfer of unentangled
 qubits \cite{gisinreview}. Later, Bru{\ss} introduced another experimentally
 feasible prepare-and-measure scheme known as the six-state scheme
 \cite{sixstate}. In her scheme, Alice randomly and independently prepares each
 qubit in one of the following six states: $|0\rangle$, $|1\rangle$, $(|0
 \rangle \pm |1\rangle)/\sqrt{2}$ and $(|0\rangle \pm i |1\rangle)/\sqrt{2}$;
 and Bob measures each of them randomly and independently in the following
 three bases: $\{ |0\rangle, |1\rangle \}$, $\{ (|0\rangle \pm |1\rangle)/
 \sqrt{2} \}$ and $\{ (|0\rangle \pm i|1\rangle)/\sqrt{2} \}$. Although the
 six-state scheme is more complex and generates a key less efficiently,
 Bru{\ss} found that it tolerates higher noise level than BB84 if Eve attacks
 each qubit individually \cite{sixstate}. In addition to qubit-based schemes
 such as BB84 and the six-state scheme, a number of QKD schemes involving
 higher dimensional as well as continuous systems have been proposed
 \cite{cont1,cont2,squeez,threestate,alph,highdim,dlevel,highdimmore}.
 Most importantly, studies showed that many schemes involving higher
 dimensional systems result in a lower fidelity of the quantum signal than
 those involving qubits under individual particle attack
 \cite{alph,highdim,dlevel,highdimmore,evethreestate}.

 Are these QKD schemes really secure? Is it really true that the six-state
 scheme tolerates higher error level than BB84? The answers to these questions
 turn out to be highly non-trivial. Recall that the all powerful Eve may choose
 to attack the transmitted qubits collectively by applying a unitary operator
 to entangle these qubits with her quantum particles. In this situation, most
 of our familiar tools such as law of large numbers and classical probability
 theory do not apply to the resultant highly entangled non-classical state.
 These make rigorous cryptanalysis of BB84 and the six-state schemes extremely
 difficult.

 In spite of these difficulties, a few air-tight security proofs against all
 possible eavesdropping attacks for BB84 and the six-state scheme have been
 discovered. Rigorous proofs for QKD schemes with better error tolerance
 capability are also found. After a few years of work, Mayers \cite{mayersjacm}
 and Biham \emph{et al.} \cite{biham} eventually proved the security of BB84
 against all kinds of attack allowed by the known laws of quantum physics. In
 particular, Mayers showed that in BB84 a provably secure key can be generated
 whenever the channel bit error rate is less than about 7\% \cite{mayersjacm}.
 (A precise definition of bit error rate can be found in
 Def.~\ref{Def:Error_Rates} in Subsection~\ref{Subsec:Error_Rates}.) Along a
 different line, Lo and Chau \cite{lochauqkdsec} proved the security of an
 entanglement-based QKD scheme that applies up to 1/3 bit error rate by means
 of a random hashing technique based on entanglement purification \cite{bdsw}.
 Their security proof is conceptually simple and appealing. Nevertheless, their
 scheme requires quantum computers and hence is not practical at this moment.
 By ingeniously combining the essence of Mayers and Lo-Chau proofs, Shor and
 Preskill gave a security proof of BB84 that applies up to 11.0\% bit error
 rate \cite{shorpre}. This is a marked improvement over the 7\% bit error
 tolerance rate in Mayers' proof. Since then, the Shor-Preskill proof became a
 blueprint for the cryptanalysis of many QKD schemes. For instance, Lo
 \cite{sixstateproof} as well as Gottesman and Lo \cite{qkd2waylocc} extended
 it to cover the six-state QKD scheme. At the same time, the work of Gottesman
 and Lo also demonstrates that careful use of local quantum operation plus two
 way classical communication (LOCC2) increases the error tolerance rate of QKD
 \cite{qkd2waylocc}. Furthermore, they found that the six-state scheme
 tolerates a higher bit error rate than BB84 because the six-state scheme gives
 better estimates for the three Pauli error rates \cite{qkd2waylocc}. In search
 of a qubit-based QKD scheme that tolerates higher bit error rate, Chau
 recently discovered an adaptive entanglement purification procedure inspired
 by the technique used by Gottesman and Lo in Ref.~\cite{qkd2waylocc}. He
 further gave a Shor-Preskill-based proof showing that this adaptive
 entanglement purification procedure allows the six-state scheme to generate a
 provably secure key up to a bit error rate of $(5-\sqrt{5})/10 \approx 27.6\%$
 \cite{sixstateexact}, making it the most error-tolerant prepare-and-measure
 scheme involving unentangled qubits to date.
 
 Unlike various qubit-based QKD schemes, a rigorous security proof against the
 most general type of eavesdropping attack on a QKD scheme involving higher
 dimensional quantum systems is lacking. Besides, the error tolerance
 capability for this kind of QKD schemes against the most general eavesdropping
 attack is virtually unexplored. In fact, almost all relevant cryptanalysis
 focus on individual particle attack; and they suggest that QKD schemes
 involving higher dimensional systems may be more error-tolerant
 \cite{alph,highdim,dlevel,evethreestate}. It is, therefore, instructive to
 give air-tight security proofs and analyze the error tolerance capability for
 this type of schemes.

 In this paper, I analyze the security and error tolerance capability of a
 prepare-and-measure QKD scheme involving the transmission of higher
 dimensional quantum systems. In fact, this scheme makes use of $N$-dimensional
 quantum states prepared and measured randomly in $(N+1)$ different bases.
 Because of the randomization of bases, the probabilities of certain kinds of
 quantum errors in the transmitted signal are correlated. This makes the error
 estimation effective and hence the error tolerance rate high. Nonetheless, the
 high error tolerance rate comes with a price, namely, that the efficiency of
 the scheme is lowered. Now, let me first begin by briefly reviewing the
 general assumptions on the capabilities of Alice, Bob and Eve together with a
 precisely stated security requirement for a general QKD scheme in
 Section~\ref{Sec:Sec_Cond}. Then, I introduce an entanglement-based QKD scheme
 involving the transmission of $N$-dimensional quantum systems where $N$ is a
 prime power in Section~\ref{Sec:EntScheme} and prove its security against the
 most general eavesdropping attack in Section~\ref{Sec:SecEntScheme}. By
 standard Shor and Preskill reduction argument, I arrive at the provably secure
 prepare-and-measure scheme in Section~\ref{Sec:ShorPre}. Since one may use $n$
 possibly entangled qubits to represent an $N$-dimensional quantum state
 whenever $N=2^n$, I obtain an unconditionally secure prepare-and-measure QKD
 scheme based on entangled qubits. This entangled-qubit-based QKD scheme offers
 a definitive advantage over all currently known unentangled-qubit-based ones
 on combating certain kind of eavesdropping strategies. More precisely, there
 is a specific eavesdropping attack that creates a bit error rate too high for
 any unentangled-qubit-based prepare-and-measure QKD scheme known to date to
 generate a provably secure key. In contrast, the same eavesdropping attack
 does not prevent this entangled-qubit-based preapre-and-measure scheme from
 producing a provably secure key whenever $N\geq 16$. But on the other hand,
 there is another specific eavesdropping attack that the entangled-qubit-based
 scheme cannot generate a provably secure key while the unentangled-qubit-based
 prepare-and-measure scheme proposed by Chau in Ref.~\cite{sixstateexact} can.
 Thus, using entangled particles as information carriers is a feasible way to
 generate a secure key under certain drastic eavesdropping attack. Lastly, I
 give a brief summary in Section~\ref{Sec:Dis}.

\section{General Features And Security Requirements For Quantum Key
 Distribution \label{Sec:Sec_Cond}}
 In QKD, we assume that Alice and Bob have access to two communication
 channels. The first one is an insecure noisy quantum channel. The other one is
 an unjammable noiseless authenticated classical channel in which everyone,
 including Eve, can listen to but cannot alter the content passing through it.
 We also assume that Alice and Bob have complete control over the apparatus in
 their own laboratories; and everything outside their laboratories except the
 unjammable classical channel may be manipulated by the all powerful Eve. We
 further make the most pessimistic assumption that Eve is capable of performing
 any operation in her controlled territory that is allowed by the known laws of
 quantum physics \cite{gottloreview,gisinreview}.

 Given an unjammable classical channel and an insecure quantum channel, a QKD
 scheme consists of three stages \cite{biasedbb84}. The first is the signal
 preparation and transmission stage where quantum signals are prepared and
 exchanged between Alice and Bob. The second is the signal quality test stage
 where a subset of the exchanged quantum signals is measured in order to
 estimate the eavesdropping rate in the quantum channel. The final phase is the
 signal privacy amplification stage where a carefully designed privacy
 amplification procedure is performed to distill out an almost perfectly secure
 key.

 No QKD scheme can be 100\% secure as Eve may be lucky enough to guess the
 preparation or measurement bases for each quantum state correctly. Hence, it
 is more reasonable to demand that the mutual information between Eve's
 measurement results after eavesdropping and the final secret key is less than
 an arbitrary but fixed small positive number. Hence I adopt the following
 definition of security.

\par\medskip
\begin{Def}[Based on Lo and Chau \cite{lochauqkdsec}]
 With the above assumptions on the unlimited computational power of Eve, a QKD
 scheme is said to be \textbf{unconditionally secure} with security parameters
 $(\epsilon_p,\epsilon_I)$ provided that whenever Eve has a cheating strategy
 that passes the signal quality control test with probability greater than
 $\epsilon_p$, the mutual information between Eve's measurement results after
 eavesdropping and the final secret key is less than $\epsilon_I$.
 \label{Def:Security}
\end{Def}

\section{An Entanglement-Based Quantum Key Distribution Scheme
\label{Sec:EntScheme}}
 In what follows, I first explicitly construct a unitary operator $T$ which
 plays a pivotal role in the design of the QKD scheme in
 Subsection~\ref{Subsec:TExist}. Then, I make use of the operator $T$ to
 construct the entanglement-based QKD scheme in
 Subsection~\ref{Subsec:EntScheme}.

\subsection{The Unitary Operator $T$ \label{Subsec:TExist}}
 In the analysis of certain quantum error correcting codes, Gottesman
 introduced a unitary operator that cyclically permutes the $\sigma_x$,
 $\sigma_y$ and $\sigma_z$ errors by conjugation \cite{QECCHammingBound}. Later
 on, Lo observed that conjugation by the same operator permutes the three bases
 used by the six-state scheme, namely, $\{ |0\rangle, |1\rangle \}$, $\{ (|0
 \rangle \pm |1\rangle)/\sqrt{2} \}$ and $\{ (|0\rangle \pm i|1\rangle)/
 \sqrt{2} \}$. He further used the permuting property of this unitary operator
 to argue that the $\sigma_x$, $\sigma_y$ and $\sigma_z$ error rates of the
 transmitted quantum signals in the six-state scheme are equal
 \cite{sixstateproof}. This is an important step in the analysis of the error
 tolerance rate of the six-state scheme as it greatly restricts the possible
 form of error in the transmitted quantum signals.

 To devise a highly error-tolerant higher dimensional QKD scheme, one naturally
 asks if it is possible to find a unitary operator $T$ that cyclically permutes
 as many types of single quantum register errors as possible by conjugation. In
 this subsection, I am going to show that such an operator $T$ indeed exists by
 explicitly writing down an expression for $T$. But before doing so, I need to
 introduce a few notations.

\par\medskip
\begin{Def}[Ashikhmin and Knill \cite{knill}]
 Suppose $a\in GF(N)$ where $N = p^n$ with $p$ being a prime. We define the
 unitary operators $X_a$ and $Z_a$ acting on an $N$-dimensional Hilbert space
 by
 \begin{equation}
  X_a |b\rangle = |a+b\rangle \label{E:DefX_a}
 \end{equation}
 and
 \begin{equation}
  Z_a |b\rangle = \chi_a (b) |b\rangle \equiv \omega_p^{\textrm{Tr} (a b)}
  |b\rangle , \label{E:DefZ_a}
 \end{equation}
 where $\chi_a$ is an additive character of the finite field $GF(N)$,
 $\omega_p$ is a primitive $p$th root of unity and $\textrm{Tr} (a) = a + a^p +
 a^{p^2} + \cdots + a^{p^{n-1}}$ is the absolute trace of $a\in GF(N)$. Note
 that, the arithmetic inside the state ket and in the exponent of $\omega_p$
 is performed in the finite field $GF(N)$. \label{Def:X_a_and_Z_a}
\end{Def}

\par\medskip
 It is easy to see from Definition~\ref{Def:X_a_and_Z_a} that $\{ X_a Z_b : a,b
 \in GF(N)\}$ spans the set of all possible linear operators for an
 $N$-dimensional quantum register over ${\mathbb C}$. Besides, $X_a$ and $Z_b$
 follow the algebra
\begin{equation}
 X_a X_b = X_b X_a = X_{a+b} , \label{E:Algebra_X_X}
\end{equation}
\begin{equation}
 Z_a Z_b = Z_b Z_a = Z_{a+b} \label{E:Algebra_Z_Z}
\end{equation}
 and
\begin{equation}
 Z_b X_a = \omega_p^{\textrm{Tr} (a b)} X_a Z_b \label{E:Algebra_X_Z}
\end{equation}
 for all $a,b\in GF(N)$, where arithmetic in the subscripts is performed in
 $GF(N)$.

 Let $T$ be a linear operator acting on an $N$-dimensional space where $N =
 p^n$ is a prime power. Inspired by the permuting property of the unitary
 operator used by Lo in the security proof of the six-state scheme
 \cite{sixstateproof}, one naturally demands that $T^{-1} X_a Z_b T =
 \omega_p^{f(a,b)} X_{a'(a,b)} Z_{b'(a,b)}$ for all $a,b\in GF(N)$. The factor
 $\omega_p^{f(a,b)} \in {\mathbb C}$ satisfying $|\omega_p^{f(a,b)}| = 1$ is
 sometimes known as the global phase because it simply multiplies a quantum
 state by a phase independent of that state. In order for $T$ to cyclically
 permute as many $X_a Z_b$'s as possible, one may demand that
\begin{equation}
 \left[ \begin{array}{c} a' \\ b' \end{array} \right]
 = \left[ \begin{array}{cc} \alpha & \beta \\ \beta & \gamma \end{array}
  \right] \,\left[ \begin{array}{c} a \\ b \end{array} \right]
 \equiv M(T) \left[ \begin{array}{c} a \\ b \end{array} \right] ,
 \label{E:Matrix}
\end{equation}
 for all $a,b \in GF(N)$, where $\alpha$, $\beta$ and $\gamma \in GF(N)$. I
 shall simply denote $M(T)$ by $M$ in this paper when the map $T$ is clearly
 known to readers.

 The phase factor $\omega_p^{f(a,b)}$ and the matrix $M(T)$ cannot be
 arbitrarily chosen. To show this, I use
 Eqs.~(\ref{E:Algebra_X_X})--(\ref{E:Matrix}) to manipulate the expression
 $X_{a+c} Z_{b+d} T$. On the one hand, it equals $\omega_p^{f(a+c,b+d)} T
 X_{(a+c)\alpha+(b+d)\beta} Z_{(a+c)\beta+(b+d)\gamma}$. On the other hand, it
 equals $\omega_p^{-\textrm{Tr}(b c)} X_a Z_b X_c Z_d T$ $=
 \omega_p^{f(c,d)-\textrm{Tr}(b c)} X_a Z_b T X_{c\alpha+d\beta} Z_{c\beta + d
 \gamma}$ $= \omega_p^{f(a,b)+f(c,d)}$ $\omega_p^{\textrm{Tr}([a\beta+b\gamma]
 [c\alpha+d\beta]-b c)}$ $T X_{(a+c)\alpha+(b+d)\beta} Z_{(a+c)\beta+(b+d)
 \gamma}$. Therefore, $T$ is well-defined if and only if the phases in the
 above two ways of expressing $X_{a+c} Z_{b+d} T$ agree for all $a,b,c,d\in
 GF(N)$.

 It is tedious but straight-forward to check that the following three
 constraints (Eqs.~(\ref{E:constraint_alpha})--(\ref{E:phase_factor})) plus the
 three phase conventions
 (Eqs.~(\ref{E:phase_convention_pgt2})--(\ref{E:phase_convention_p2b})) make
 the expressions in the above paragraph consistent and hence the linear map $T$
 well-defined:
\begin{equation}
 \alpha \gamma - \beta^2 = 1 , \label{E:constraint_alpha}
\end{equation}
\begin{equation}
 X_a Z_b T = \omega_p^{f(a,b)} T X_{a\alpha + b\beta} Z_{a\beta + b\gamma}
 \label{E:XZT}
\end{equation}
 and
\begin{eqnarray}
 f(a,b) & = & \frac{1}{2} \textrm{Tr} (\beta [ a^2 \alpha + b^2 \gamma]) +
 \textrm{Tr} (a b \beta^2 + \nonumber \\
 & & ~~\delta_{p2} \beta \sum_{i>j} g_i g_j [a_i a_j \alpha + b_i b_j \gamma ])
 \label{E:phase_factor}
\end{eqnarray}
 for all $a,b\in GF(N)$. Note that in Eq.~(\ref{E:phase_factor}), $a =
 \sum_{i=1}^n a_i g_i$ and $b = \sum_{i=1}^n b_i g_i$ where $\{ g_1, g_2,
 \ldots , g_n \}$ is a fixed basis of $GF(N)$ over the field $GF(p)$ and $a_i,
 b_i \in GF(p)$. Moreover, $\delta_{p2}$ in the above equation is the Kronecker
 delta.

 Two important remarks are in place. First, when $p>2$ and hence $N$ is odd,
 $2$ is invertible in $GF(N)$. Consequently, global phase $\omega_p^{f(a,b)}$
 may be chosen from $p$th roots of unity. Following this convention, I demand
\begin{equation}
 f(a,b) \in \mathbb{Z}/p{\mathbb Z} \textrm{~for any~} a,b\in GF(N)
 \textrm{~if~} 2\not| N . \label{E:phase_convention_pgt2}
\end{equation}
 In contrast, when $p=2$ and hence $N$ is even, $2$ is not invertible in
 $GF(N)$. In this case $f(a,b)$ may be integral or half-integral. Consequently,
 $\omega_p^{f(a,b)} \in \{ \pm 1,\pm i \}$. In this case, I use the convention
 that
\begin{equation}
 \omega_2^{\textrm{Tr} (\alpha\beta a_j^2 g_j^2)/2} = \left\{ \begin{array}{ll}
 1 & \textrm{if~} \textrm{Tr} (\alpha\beta a_j^2 g_j^2) = 0, \\ i &
 \textrm{if~} \textrm{Tr} (\alpha\beta a_j^2 g_j^2) = 1, \end{array} \right.
 \label{E:phase_convention_p2a}
\end{equation}
 and
\begin{equation}
 \omega_2^{\textrm{Tr} (\beta\gamma b_j^2 g_j^2)/2} = \left\{ \begin{array}{ll}
 1 & \textrm{if~} \textrm{Tr} (\beta\gamma b_j^2 g_j^2) = 0, \\ i &
 \textrm{if~} \textrm{Tr} (\beta\gamma b_j^2 g_j^2) = 1, \end{array} \right.
 \label{E:phase_convention_p2b}
\end{equation}
 for all $a_j,b_j\in GF(p)$, where $j=1,2,\ldots ,n$.

 The second remark concerns the reason why we have the last term in
 Eq.~(\ref{E:phase_factor}). Recall that the identity $\textrm{Tr} (a_i^2 +
 a_j^2)/2 + \textrm{Tr} (a_i a_j) = \textrm{Tr} ([a_i + a_j]^2)/2$ holds only
 for $p>2$. In contrast, $\textrm{Tr} (a_i^2 + a_j^2) = \textrm{Tr} ([a_i +
 a_j]^2)$ for $p=2$. So, I cannot use the first identity to absorb the last
 term in Eq.~(\ref{E:phase_factor}) into the first term when $p=2$.

\par\medskip
\begin{Lem}
 A linear operator $T$ obeying
 Eqs.~(\ref{E:constraint_alpha})--(\ref{E:phase_convention_p2b}) is unitary
 after a proper scaling. Specifically, $T$ is unitary if and only if its
 operator norm satisfies $\| T \| = 1$. \label{Lem:Unitary_Condition}
\end{Lem}
\begin{proof}
 I only need to show that $\| T \| = 1$ is a sufficient condition as this
 condition is clearly necessary.
 Eqs.~(\ref{E:constraint_alpha})--(\ref{E:phase_convention_p2b}) lead to $X_a
 Z_b T T^\dagger =$ $\omega_p^{f(a,b)} T X_{a\alpha + b\beta} Z_{a \beta + b
 \gamma} T^\dagger =$ $\omega_p^{f(a,b) - \textrm{Tr} ([a\alpha + b\beta ][a
 \beta + b\gamma])} T Z_{-a\beta-b\gamma}^\dagger X_{-a\alpha-b\beta}^\dagger
 T^\dagger =$ $\omega_p^{f(a,b)+f(-a,-b) - \textrm{Tr} (\beta [a^2 \alpha + b^2
 \gamma]) - 2 \textrm{Tr} (a b \beta^2)} T T^\dagger X_a Z_b =$ $T T^\dagger
 X_a Z_b$ for all $a,b\in GF(N)$. By the same argument, $X_a Z_b T^\dagger T =
 T^\dagger T X_a Z_b$ for all $a,b\in GF(N)$. Since $T$ acts on a finite
 dimensional Hilbert space and $\{ X_a Z_b : a,b\in GF(N) \}$ spans the set of
 all linear operators on that Hilbert space, $T T^\dagger$ and $T^\dagger T$
 are constant multiples of the identity operator. Therefore, $\| T \| = 1$
 implies $T T^\dagger = I = T^\dagger T$. Hence, $T$ is unitary.
\end{proof}

\par\medskip
 In order to fully utilize the error tolerance capability of an $N$-dimensional
 QKD scheme, $T$ should satisfy one more constraint, namely, the order of $T$
 must be as large as possible. The theorem below gives us an attainable upper
 bound for the order of $T$.

\par\medskip
\begin{Thrm}
 There exists a unitary operator $T$ satisfying the constraints
 Eqs.~(\ref{E:constraint_alpha})--(\ref{E:phase_factor}), the phase conventions
 stated in
 Eqs.~(\ref{E:phase_convention_pgt2})--(\ref{E:phase_convention_p2b})
 as well as the condition that $I, T, T^2, \ldots , T^N$ are distinct operators
 up to a global phase. (That is, for all $0\leq i < j \leq N$ and $\theta\in
 {\mathbb R}$, $T^i \neq e^{i\theta} T^j$.) Furthermore, the order of $T$
 up to a global phase satisfying
 Eqs.~(\ref{E:constraint_alpha})--(\ref{E:phase_convention_p2b}) is at most
 $(N+1)$. Suppose further that $\{ g_1, g_2, \ldots , g_n \}$ is a fixed basis
 of $GF(N)$ over $GF(p)$, then $T$ is given by
\begin{equation}
 T = \frac{e^{i\theta}}{N} \sum_{a,b\in GF(N)} \omega_p^{\textrm{Tr} (\varphi_1
 (a,b)) - \frac{1}{2} \textrm{Tr} (\varphi_2 (a,b))} X_a Z_b
 \label{E:Explicit_T}
\end{equation}
 for some $\theta\in {\mathbb R}$, where
\begin{eqnarray}
 & & \varphi_1 (a,b) \nonumber \\
 & = & \frac{1}{(2-\alpha-\gamma)^2} \{ \beta^3 (\gamma -1) a^2 -(\gamma-1) [
 (\alpha-1)^2 + \nonumber \\
 & & ~~\beta^2 (2\alpha -1)] a b + \beta [ \alpha \gamma (\alpha-1) + \gamma -
 1] b^2 \} + \nonumber \\
 & & ~~\delta_{p2} \beta \sum_{i>j} g_i g_j (\tilde{a}_i \tilde{a}_j \alpha +
 \tilde{b}_i \tilde{b}_j \gamma )
 \label{E:Explicit_varphi1}
\end{eqnarray}
 and
\begin{eqnarray}
 & & \varphi_2 (a,b) \nonumber \\
 & = & \frac{\beta}{(2-\alpha-\gamma)^2} \,[ (\alpha + \gamma - 2\alpha \gamma)
 (a^2 + 2\beta a b + b^2) + \nonumber \\
 & & ~~2\beta^2 (\gamma a^2 + \alpha b^2) ] .
 \label{E:Explicit_varphi2}
\end{eqnarray}
 Note that all the arithmetic in the above two equations are performed in the
 finite field $GF(N)$. Besides, in Eq.~(\ref{E:Explicit_varphi1}),
 $\tilde{a}_i,\tilde{b}_i \in GF(p)$ are the unique solutions of the equations
\begin{equation}
 \sum_{i=1}^n \tilde{a}_i g_i = \frac{(\gamma-1)a-\beta b}{2-\alpha-\gamma}
 \label{E:solution_tilde_a}
\end{equation}
 and
\begin{equation}
 \sum_{i=1}^n \tilde{b}_i g_i = \frac{(\alpha-1)b-\beta a}{2-\alpha-\gamma} .
 \label{E:solution_tilde_b}
\end{equation}
 \label{Thrm:Good_T_Exist}
\end{Thrm}
\begin{proof}
 From Eqs.~(\ref{E:Matrix}) and~(\ref{E:XZT}), I know that the order of $T$ up
 to a global phase is equal to the order of $M\equiv M(T)$. Combining with
 Eq.~(\ref{E:constraint_alpha}), the characteristic equation of $M$ is
 $\textrm{Char} (M) = \lambda^2 - (\alpha + \gamma) \lambda + 1$. If
 $\textrm{Char} (M)$ is reducible in $GF(N)$, the order of $M$ and hence also
 the order of $T$ up to a global phase are at most $(N-1)$. So, to construct
 $T$ with a larger order, I must look for $\textrm{Char} (M)$ that is
 irreducible in $GF(N)$. Nevertheless, a degree two irreducible polynomial over
 $GF(N)$ splits in $GF(N^2)$. Since the constant term of $\textrm{Char} (M)$ is
 1, the roots of $\textrm{Char} (M) = 0$ over $GF(N^2)$ can be written as $\xi$
 and $\xi^{-1}$ respectively. Since $\alpha + \gamma \in GF(N)$, I conclude
 that $\xi + \xi^{-1} = (\xi + \xi^{-1})^N = \xi^N + \xi^{-N}$. Therefore,
 $(\xi^{N+1} -1)(\xi^{N-1} -1) = 0$. However, $\xi\not\in GF(N)$ and hence
 $\xi^{N+1} = 1$. In other words, the order of the irreducible polynomial
 $\textrm{Char} (M)$ and hence the order of $T$ up to a global phase both
 divide $(N+1)$. More importantly, since $N \not\equiv 1 \bmod (N+1)$ and $N^2
 \equiv 1 \bmod (N+1)$, Theorem~3.5 in Ref.~\cite{finitefieldbook} assures the
 existence of an order $(N+1)$ irreducible polynomial in the form $\lambda^2 +
 c \lambda + 1$ over $GF(N)$. (Actually, Theorem~3.5 in
 Ref.~\cite{finitefieldbook} implies that $\lambda^2 + c\lambda + 1$ is
 irreducible over $GF(N)$ if and only if it is equal to $(\lambda + \xi)
 (\lambda + \xi^{-1})$ for $\xi \in GF(N^2) \backslash GF(N)$ with $\xi^{N+1} =
 1$. Hence, such irreducible polynomials can be found efficiently.)

 It remains to show that there exists $T$ whose order of the corresponding
 characteristic polynomial $\textrm{Char} (M(T))$ equals $(N+1)$. I divide the
 proof into two cases.

 Case 1:~$p = 2$ or $p \equiv 1 \bmod 4$ where $N = p^n$. In this case, I
 simply pick $\alpha = 0$, $\gamma = -c$ and $\beta = (-1)^{1/2}$. (Such a
 $\beta \in GF(N)$ exists because $x^2 \equiv -1 \bmod p$ is solvable when $p =
 2$ or $p$ is a prime satisfying $p \equiv 1 \bmod 4$.) Then, it is easy to
 check that Eq.~(\ref{E:constraint_alpha}) is satisfied and hence $T$ exists.

 Case 2:~$p > 2$. In this case, I pick $\alpha = 1$, $\gamma = -c-1$. In this
 way, $\beta^2 = -c-2 = \xi + \xi^{-1} -2 = -(\xi - 1) (\xi^{-1} -1) = (\xi
 -1)^2 \xi^{-1}$. Hence, I choose $\beta = (\xi - 1) \xi^{-1/2} = \xi^{1/2} -
 \xi^{-1/2}$. ($\xi^{1/2}$ exists since $p$ is an odd prime and $\xi^{N+1} =
 1$ so that $\xi = \kappa^{N-1}$ where $\kappa$ is a primitive element in
 $GF(N^2)$. Moreover, $\beta\in GF(N)$ since $(\xi^{1/2} - \xi^{-1/2})^N =
 \xi^{N/2} - \xi^{-N/2} = -\xi^{-1/2} + \xi^{1/2}$.)

 Now, I am ready to explicitly construct $T$. To do so, I write $T = \sum_{a,b
 \in GF(N)} \Lambda_{ab} X_a Z_b$ for some $\Lambda_{ab} \in {\mathbb C}$. From
 Eq.~(\ref{E:XZT}), I conclude that
\begin{eqnarray}
 \Lambda_{ij} & = & \omega_p^{f(a,b) + \textrm{Tr} ([a\alpha + b\beta] \{j - a
 \beta - b [ \gamma -1] \} - b i)} \times \nonumber \\
 & & ~~\Lambda_{i-a(\alpha-1)-b\beta,j-a\beta-b(\gamma-1)} \label{E:T_relation}
\end{eqnarray}
 for all $a,b,i,j\in GF(N)$. Since the order of $T$ is greater than 1,
 $M(T)-I = \left[ \begin{array}{cc} \alpha-1 & \beta \\ \beta & \gamma-1
 \end{array} \right]$ is invertible. Hence, I can choose suitable $a = a(i,j)$
 and $b = b(i,j)$ in Eq.~(\ref{E:T_relation}) to relate every $\Lambda_{ij}$ to
 $\Lambda_{00}$. In this way, I conclude that every $\Lambda_{ij}$ is
 proportional to $\Lambda_{00}$. Besides, all $|\Lambda_{ij}|$'s are equal.
 Consequently, the unitarity of $T$ implies that $|\Lambda_{00}| = 1/N$. By
 explicitly substituting $a,b$ into Eq.~(\ref{E:T_relation}) and after a
 tedious but straight-forward calculation, I arrive at
 Eqs.~(\ref{E:Explicit_T})--(\ref{E:solution_tilde_b}).
\end{proof}

\par\medskip
 The explicit construction of the operator $T$ in the above proof also shows
 that once the $2\times 2$ matrix $M(T)$ and the primitive root $\omega_p$ are
 fixed, $T$ is uniquely determined up to a global phase and a convention for
 $\omega_p^{f(a,b)}$.

 For illustration purpose, the choices of $M(T)$'s and hence the unitary
 operators $T$'s for $N = 2,3,4$ computed by
 Eqs.~(\ref{E:phase_convention_pgt2})--(\ref{E:solution_tilde_b}) are tabulated
 in Table~\ref{T:T}. Incidentally, the unitary operator $T$ listed in
 Table~\ref{T:T} for $N=2$ is, up to a global phase, the same as the one used
 by Lo in his security proof of the six-state scheme in
 Ref.~\cite{sixstateproof}.

\begin{table}[t]
\begin{center}
\begin{tabular}{|c|c|l|}
 \hline
 $N$ & $M(T)$ & $T$ \\ \hline
 2 & $\left[ \begin{array}{cc} 0 & 1 \\ 1 & 1 \end{array} \right]$ & 
  $\displaystyle \frac{1}{2} \left( I - i X_1 - i Z_1 + X_1 Z_1 \right)$ \\
 \hline
 3 & $\left[ \begin{array}{cc} 1 & 1 \\ 1 & 2 \end{array} \right]$ &
  $\displaystyle \frac{1}{3} \sum_{i,j=0}^2 \omega_3^{2\delta_{i0}+\delta_{j0}}
  X_i Z_j$ \\ \hline
 4 & $\left[ \begin{array}{cc} 0 & 1 \\ 1 & \omega \end{array} \right]$ &
  $\displaystyle \frac{1}{4} \!\!\sum_{i,j\in GF(4)} \!\!\!(-1)^{-\textrm{Tr}
  (\omega [i+j])/2 + \textrm{Tr} (i+j) + \delta_{i\!+\!j\!-\!1}} X_i Z_j$ \\
 \hline
\end{tabular}
\end{center}
\caption{The choices of $T$ and $M(T)$ for $N=2$, $3$ and $4$. Note that
 $\omega\in GF(4)$ satisfies $\omega^2 + \omega + 1 = 0$ and I have used $\{ 1,
 \omega \}$ as the basis of $GF(4)$ over $GF(2)$ when constructing $T$ for $N
 = 4$.}
\label{T:T}
\end{table}

 Now, I report several important properties of $T$ and $M(T)$ that will be used
 in the security proof of this QKD scheme in Section~\ref{Sec:SecEntScheme}.

\par\medskip
\begin{Lem}
 Suppose the order of $M(T)$ equals $(N+1)$, then $M(T)^k$ is in the form $a I$
 for some $a\in GF(N)$ if and only if (1)~$p = 2$ and $(N+1) | k$; or (2)~$p >
 2$ and $[(N+1)/2] | k$. In fact, if $p>2$, $M(T)^{(N+1)/2} = -I$.
 \label{Lem:Order}
\end{Lem}
\begin{proof}
 Since $\textrm{Char} (M(T)) = \lambda^2 + c\lambda + 1$, $M(T)$ can be written
 in the form $P^{-1} D P$ where $D = \textrm{diag} (\xi , \xi^{-1})$ where $\xi
 \in GF(N^2)$ and $\xi^{N+1} = 1$. Hence $M(T)^k = a I$ if and only if
 $\xi^{2k} = 1$. If $p = 2$, $\xi^{2k} = 1 \Leftrightarrow \xi^k = 1
 \Leftrightarrow (N+1) | k$. And if $p > 2$, $\xi^{2k} = 1 \Leftrightarrow
 \xi^k = \pm 1 \Leftrightarrow [(N+1)/2] | k$. Moreover, $\xi^k = -1$ if and
 only if $k = [(N+1)/2] \bmod (N+1)$.
\end{proof}

\par\medskip
\begin{Cor}
 The period of the sequence $\{ T^{-k} X_a Z_b T^k: k\in {\mathbb N} \}$ up to
 global phases equals $(N+1)$ whenever $a,b\in GF(N)$ are not all zero.
 Furthermore, if $p=2$, there is exactly one $0\leq k\leq N$ with $T^{-k} X_a
 Z_b T^k = \Lambda Z_c$ for some $\Lambda\in {\mathbb C}$ and $c\in GF(N)$. If
 $p>2$, either $T^{-k} X_a Z_b T^k \neq \Lambda Z_c$ for all $k$ or there are
 two distinct $0\leq k,k'\leq N$ with $T^{-k} X_a Z_b T^k = \Lambda Z_c$ and
 $T^{-k'} X_a Z_b T^{k'} = \Lambda' Z_{c'}$ for $\Lambda,\Lambda' \in
 {\mathbb C}$ and $c\neq c' \in GF(N)$. \label{Cor:Sequence}
\end{Cor}
\begin{proof}
 Direct application of Lemma~\ref{Lem:Order}.
\end{proof}

\par\medskip
\begin{Def}
 $T$ defines an equivalent relationship for $GF(N)^2$ by $(a,b) \sim (a',b')$
 if there exists $i\in {\mathbb N}$ and $\Lambda\in {\mathbb C}\backslash \{ 0
 \}$ such that $T^{-i} X_a Z_b T^i = \Lambda X_{a'} Z_{b'}$. I denote elements
 in the corresponding equivalent class by $(a,b)/\sim$. \label{Def:Equivalent}
\end{Def}

\par\medskip
\begin{Cor}
 There are $N$ elements in the equivalent class $GF(N)^2/\sim$. Besides, $|
 (a,b) / \sim | = N+1$ if $(a,b) \neq (0,0)$. For every $a\in GF(N)$, there
 exists at most two distinct $b,b'\in GF(N)$ such that $(a,b)\sim (a,b')$.
 Furthermore, if $p>2$, $b\neq b'$ and $c\neq 0$, then $(0,c) \sim (a,b) \sim
 (a,b') \Rightarrow a=0$ if and only if $N=3$. If $p=2$, $(0,b) \sim (0,b')$
 implies $b=b'$. In addition, suppose that $p=2$ and $a\neq 0$. Then, for any
 $b\in GF(N)$, there exists $c=c(b)$ such that $(0,a) \sim (b,c)$. In summary,
 $GF(N)^2/\sim = \{ (0,a)/\sim : a\in GF(N) \}$ if $p=2$. On the other hand, if
 $p>2$, there are $(N-1)/2$ elements of $GF(N)^2/\sim$ each containing two
 distinct elements in the form $(0,b)$. \label{Cor:EquivalentClass}
\end{Cor}
\begin{proof}
 By writing 
\begin{eqnarray}
 & & M(T) \nonumber \\
 & = & P^{-1} \left[ \begin{array}{cc} \xi & 0 \\ 0 & \xi^{-1} \end{array}
  \right] P \nonumber \\
 & \equiv & \left[ \begin{array}{cc} \beta & \xi - \alpha \\ \beta & \xi^{-1} -
  \alpha \end{array} \right]^{-1} \left[ \begin{array}{cc} \xi & 0 \\ 0 &
  \xi^{-1} \end{array} \right] \left[ \begin{array}{cc} \beta & \xi - \alpha \\
  \beta & \xi^{-1} - \alpha \end{array} \right] , \label{E:def_diag}
\end{eqnarray}
 then $(a,b) \sim (a,b')$ if and only if there exists $k$ such that
\begin{equation}
 \left[ \begin{array}{cc} \xi^k & 0 \\ 0 & \xi^{-k} \end{array} \right] P
 \left[ \begin{array}{c} a \\ b \end{array} \right] = P \left[ \begin{array}{c}
 a \\ b' \end{array} \right] ~. \label{E:preparation}
\end{equation}
 By eliminating $k$ from the above equation, I obtain a quadratic equation
 involving variables $a,b$ and $b'$. Thus, for a given $a,b$, there are at most
 two distinct $b'$ satisfying Eq.~(\ref{E:preparation}). Hence, for every $a\in
 GF(N)$, there are at most two distinct $b,b'\in GF(N)$ with $(a,b)\sim
 (a,b')$.
 
 Now suppose $p>2$, $b\neq b'$ and $c\neq 0$. If $(0,c)\sim (a,b)\sim (a,b')$,
 there exist two distinct integers $k,k'\in [0,N]$ such that $M^k [0~c]^T =
 [a~b]^T$ and $M^{k'} [0~c]^T = [a~b']^T$. Using Eq.~(\ref{E:def_diag}) to
 equate the first rows of the above two equations, I obtain $\xi^k - \xi^{-k} =
 \xi^{k'} - \xi^{-k'}$. The solution of this equation is $\xi^k = \xi^{k'}$ or
 $\xi^{k+k'} = -1$. Since $p>2$, Lemma~\ref{Lem:Order} demands that $k=k' \bmod
 (N+1)$ or $k+k' = [(N+1)/2] \bmod (N+1)$. As $N$ is odd, there are at most two
 solutions for $2k = [(N+1)/2] \bmod (N+1)$. Thus, provided that $N>3$, there
 exist more than two pairs of $(k,k')$ such that $k\neq k'$ and $k+k' =
 [(N+1)/2] \bmod (N+1)$. Hence, there exist $b\neq b'$ such that $(0,c) \sim
 (a,b) \sim (a,b')$ for $a\neq 0$. In contrast, if $N=3$, $(0,2)$ and $(2,0)$
 are the only two pairs of $(k,k')$ satisfying $k\neq k'$ and $k+k' = 2 \bmod
 4$. From Lemma~\ref{Lem:Order}, $M^2 = -I$ when $N=3$. Hence, $(a,b,b')$
 equals $(0,1,2)$ or $(0,2,1)$. Therefore, $(0,c) \sim (a,b) \sim (a,b')
 \Rightarrow a=0$.

 The remaining assertions then follow directly from
 Corollary~\ref{Cor:Sequence}. 
\end{proof}

\subsection{An Entanglement-Based QKD Scheme \label{Subsec:EntScheme}}
 Let $N$ be a prime power and $T$ be the order $(N+1)$ unitary operator
 described in Theorem~\ref{Thrm:Good_T_Exist} in
 Subsection~\ref{Subsec:TExist}. Then, the QKD scheme goes as follows.

\par\medskip
\begin{enumerate}[\underline{Entanglement-based QKD Scheme~A}]
 \item Alice prepares $L\gg 1$ quantum particle pairs in the state
  $\sum_{i\in GF(N)} |ii\rangle / \sqrt{N}$. She applies one of the following
  unitary transformation to the second particle in each pair randomly and
  independently: $I, T, T^2, \ldots , T^N$. For every pair of particles, Alice
  keeps the first one and sends the second one to Bob. He acknowledges the
  reception of these particles and then applies one of the following to each
  received particle randomly and independently: $I^{-1}, T^{-1}, T^{-2},\ldots,
  T^{-N}$. Now, Alice and Bob publicly reveal their unitary transformations
  applied to each particle. A shared pair is then kept and is said to be in the
  set $S_i$ if Alice and Bob have applied $T^i$ and $T^{-i}$ to the second
  particle of the shared pair respectively. Thus in the absence of noise and
  Eve, each pair of shared particles kept by Alice and Bob should be in the
  state $\sum_{i\in GF(N)} |ii\rangle /\sqrt{N}$. \label{Ent:Prepare}
 \item Alice and Bob estimate the (quantum) channel error rate by sacrificing a
  few particle pairs. Specifically, they randomly pick $\mbox{O} ([N+1]^2 \log
  \{ [N+1]/\epsilon \} / \delta^2 N^2)$ pairs from each of the $(N+1)$ sets
  $S_i$ and measure each particle of the pair in the $\{ |0\rangle, |1\rangle,
  \cdots ,|N-1\rangle \}$ basis, namely the standard basis. They publicly
  announce and compare their measurement results. In this way, they know the
  estimated channel error rate within standard deviation $\delta$ with
  probability at least $(1-\epsilon)$. (Detail proof of this claim can be found
  in Ref.~\cite{biasedbb84}. A brief outline of the proof will also be given in
  Subsection~\ref{Subsec:Sampling} for handy reference.) If the channel error
  rate is too high, they abort the scheme and start all over again.
  \label{Ent:QCTest}
  \item Alice and Bob perform the following privacy amplification procedure.
   (Readers will find out in Section~\ref{Sec:SecEntScheme} that
   step~\ref{Ent:PriAmp_SpinCorr} below reduces errors in the form $X_a Z_b$
   with $a\neq 0$ at the expense of increasing errors in the form $Z_c$ with $c
   \neq 0$. In contrast, step~\ref{Ent:PriAmp_PhaseCorr} below reduces errors
   in the form $X_a Z_b$ with $b\neq 0$ at the expense of increasing errors in
   the form $X_c$ with $c\neq 0$. Most vitally, applying
   steps~\ref{Ent:PriAmp_SpinCorr} and~\ref{Ent:PriAmp_PhaseCorr} in turn is an
   effective way to reduce all kinds of errors.)
   \begin{enumerate}
    \item Alice and Bob apply the entanglement purification procedure by two
     way classical communication (LOCC2 EP) similar to the ones reported in
     Refs.~\cite{bdsw,genpur}. Specifically, Alice and Bob randomly group their
     remaining quantum particles in tetrads; and each tetrad consists of two
     pairs shared between Alice and Bob in Step~\ref{Ent:Prepare}. Alice
     randomly picks one of the two particles in her share of each tetrad as the
     control register and the other as the target. She applies the following
     unitary operation to the control and target registers:
     \begin{equation}
      |i\rangle_\textrm{control} \otimes |j\rangle_\textrm{target} \longmapsto
      |i\rangle_\textrm{control} \otimes |j-i\rangle_\textrm{target} ,
      \label{E:GenCNOT}
     \end{equation}
     where the subtraction is performed in the finite field $GF(N)$. Bob
     applies the same unitary transformation to his corresponding share of
     particles in the tetrad. Then, they publicly announce their measurement
     results of their target registers in the standard basis. They keep their
     control registers only when the measurement results of their corresponding
     target registers agree. They repeat the above LOCC2 EP procedure until
     there is an integer $r>0$ such that a single application of
     step~\ref{Ent:PriAmp}b will bring the quantum channel error rate of the
     resultant particles down to less than $\epsilon_I / \ell^2$ for an
     arbitrary but fixed security parameter $\epsilon_I > 0$, where $r \ell$ is
     the number of remaining pairs they shared currently. They abort the scheme
     either when $r$ is greater than the number of remaining quantum pairs they
     possess or when they have used up all their quantum particles in this
     procedure. \label{Ent:PriAmp_SpinCorr}
    \item They apply the majority vote phase error correction (PEC) procedure
     introduced by Gottesman and Lo \cite{qkd2waylocc}. Specifically, Alice and
     Bob randomly divide the resultant particles into sets each containing $r$
     pairs of particles shared between Alice and Bob. Alice and Bob separately
     apply the $[r,1,r]_N$ phase error correction procedure to their
     corresponding shares of $r$ particles in each set and retain their phase
     error corrected quantum particles. At this point, Alice and Bob should
     share $\ell$ almost perfect pairs $\sum_{i\in GF(N)} |ii\rangle /
     \sqrt{N}$ with fidelity at least $(1-\epsilon_I /\ell)$. By measuring
     their shared pairs in the standard basis, Alice and Bob obtain their
     common key. More importantly, Eve's information on this common key is less
     than the security parameter $\epsilon_I$. (Proof of this claim can be
     found in Theorem~\ref{Thrm:Uncond_Sec_A} in Subsection~\ref{Subsec:PA}
     below.) \label{Ent:PriAmp_PhaseCorr}
   \end{enumerate}
   \label{Ent:PriAmp}
\end{enumerate}

\par\medskip
 Note that when $N=2$, Scheme~A is a variation of the six-state scheme
 introduced by Chau in Ref.~\cite{sixstateexact}. The key difference is that
 the present one does not make use of Calderbank-Shor-Steane quantum code after
 PEC while the former one does.

\section{Cryptanalysis Of The Entanglement-Based Quantum Key Distribution
 Scheme \label{Sec:SecEntScheme}}
 In this section, I am going to report a detail unconditional security proof of
 Scheme~A in the limit of large number of quantum particle $L$ transmitted. I
 will also investigate the maximum error tolerance rate for Scheme~A against
 the most general type of eavesdropping attack allowed by the laws of quantum
 physics. With suitable modifications, the security proof reported here can be
 extended to the case of a small finite $L$. Nevertheless, working in the limit
 of large $L$ makes the asymptotic error tolerance rate analysis easier.

 Before carrying out the cryptanalysis, I will first define various error rate
 measures and discuss how to fairly compare error tolerance capabilities
 between different QKD schemes in Subsection~\ref{Subsec:Error_Rates}. Then, I
 will briefly explain why a reliable upper bound of the channel error can be
 obtained by randomly testing only a small subset of quantum particles in
 step~\ref{Ent:QCTest} of Scheme~A in Subsection~\ref{Subsec:Sampling}.
 Finally, I will prove the security of the privacy amplification procedure in
 step~\ref{Ent:PriAmp} of Scheme~A and analyze its error tolerance rate in
 Subsection~\ref{Subsec:PA}. This will complete the proof of unconditional
 security for entanglement-based Scheme~A.

\subsection{Fair Comparison Of Error Tolerance Capability And Various Measures
 Of Error Rates \label{Subsec:Error_Rates}}
\par\medskip
\begin{Def}
 Recall that Alice prepares $L$ particle pairs each in the state $\sum_{i\in
 GF(N)} |ii\rangle/\sqrt{N}$ and randomly applies powers of $T$ to each pair.
 Denote the resultant (pure) state of the pairs by $\bigotimes_{j=1}^L |\phi_j
 \rangle$. Then, she sends one particle in each pair through an insecure
 quantum channel to Bob; and upon reception, Bob randomly applies powers of $T$
 to his share of the pair. The {\bf channel quantum error rate} in this
 situation is defined as the marginal error rate of the measurement results
 when Alice and Bob were going to make an hypothetical measurement on the $j$th
 shared quantum particle pair in the basis $\{ X_a Z_b \otimes I |\phi_j\rangle
 : a,b\in GF(N) \}$ for all $j$. In other words, the channel quantum error rate
 equals $1/L$ times the expectation value of the cardinality of the set $\{ j :
 \textrm{hypothetical measurement of the $j$th pair equals } X_a Z_b \otimes I
 |\phi_j\rangle \textrm{ with } (a,b)\neq (0,0) \}$. The {\bf channel standard
 basis measurement error rate} is defined as $1/L$ times the expectation value
 of the cardinality of the set $\{ j : \textrm{hypothetical measurement of the
 $j$th pair equals } X_a Z_b \otimes I |\phi_j\rangle \textrm{ with } a\neq 0
 \}$. The next two definitions concern only those quantum particle pairs
 retained by Alice and Bob in $\bigcup_i S_i$. (That is, those Alice and Bob
 have applied $T^j$ and $T^{-j}$ to the second particle of the shared pair for
 some $j$ respectively.) In the absence of noise and eavesdropper, all such
 particle pairs should be in the state $\sum_{i\in GF(N)} |ii\rangle/\sqrt{N}$.
 The {\bf signal quantum error rate} (or quantum error rate (QER) for short) in
 this situation is defined as the expectation value of the proportion of
 particle pairs in $\bigcup_i S_i$ whose measurement result in the basis $\{
 \sum_{i\in GF(N)} |i\rangle \otimes X_a Z_b |i\rangle/\sqrt{N}: a,b\in GF(N)
 \}$ equals $\sum_{i\in GF(N)} |i\rangle \otimes X_a Z_b |i\rangle/\sqrt{N}$
 for some $(a,b) \neq (0,0)$. The {\bf signal standard basis measurement error
 rate} (or standard basis measurement error rate (SBMER) for short) is defined
 as the expectation value of the proportion of particle pairs in $\bigcup_i
 S_i$ whose measurement result in the basis $\{ \sum_{i\in GF(N)} |i\rangle
 \otimes X_a Z_b |i\rangle/\sqrt{N} : a,b\in GF(N) \}$ equals $\sum_{i\in
 GF(N)} |i\rangle \otimes X_a Z_b |i\rangle/\sqrt{N}$ for some $a\neq 0$. In
 other words, SBMER measures the apparent error rate of the signal when Alice
 and Bob measure their shares of particles in the standard basis. In the
 special case of $N = 2^n$, any standard basis measurement result can be
 bijectively mapped to a $n$-bit string. Thus, it makes sense to define the
 {\bf signal bit error rate} (or bit error rate (BER) for short) as the
 marginal error rate of resultant $n$-bit string upon standard basis
 measurement of the signal at the end of the signal preparation and
 transmission stage. \label{Def:Error_Rates}
\end{Def}

\par\medskip
 Three important remarks are in place. First, SBMERs and BERs for QKD schemes
 using quantum particles of different dimensions as information carriers should
 \emph{never} be compared directly. This is because the quantum communication
 channels used are different. In addition, the same eavesdropping strategy may
 lead to different error rates
 \cite{alph,highdim,dlevel,highdimmore,evethreestate}. It appears that the only
 sensible situation to meaningfully compare the error tolerance capabilities of
 two QKD schemes is when the schemes are using the same quantum communication
 channel and are subjected to the same eavesdropping attack. Specifically,
 suppose Alice reversibly maps every $p^n$-dimensional quantum state used in
 Scheme~A into $n$ possibly entangled $p$-dimensional quantum particles and
 sends them through an insecure $p$-dimensional quantum particle communication
 channel to Bob. Moreover, since we assume that Alice and Bob do not have
 quantum storage capability, it is reasonable to regard Alice to send every
 packet of $n$ possibly entangled $p$-dimensional quantum particles
 consecutively. In this way, Scheme~A becomes an entangled-particle-based QKD
 scheme. More importantly, Eve may apply the same eavesdropping attack on the
 insecure $p$-dimensional quantum particle channel used by Alice and Bob
 irrespective of $n$. In this way, I can fairly compare the error tolerance
 capability between two entangled-particle-based QKD schemes derived from
 Scheme~A using $p^n$- and $p^{n'}$-dimensional particles respectively against
 any eavesdropping attack on the $p$-dimensional quantum particle channel.

 Second, the BER defined above for $N=2^n$ with $n>1$ depends on the bijection
 used. Fortunately, a useful lower bound on the BER can be found amongst all
 bijections immediately before Eq.~(\ref{E:BER_QER}) in
 Subsection~\ref{Subsec:PA}.

 Third, since quantum errors in the form $X_a Z_b$ with $(a,b)\neq (0,0)$
 permute under the conjugation by powers of $T$, the channel quantum error rate
 is equal to the QER of the signal. Roughly speaking, QER refers to the rate of
 any quantum error (phase shift and/or spin flip) occurring in the pair
 $\sum_{i\in GF(N)} |ii\rangle / \sqrt{N}$ shared by Alice and Bob. In
 contrast, due to the permutation of quantum errors by powers of $T$, the
 channel standard basis measurement error rate does not equal to the SBMER in
 general.

\subsection{Reliability On The Error Rate Estimation \label{Subsec:Sampling}}
 In Scheme~A, Alice and Bob keep only those particle pairs that are believed to
 be in the state $\sum_{i\in GF(N)} |ii\rangle / \sqrt{N}$ at the end of
 step~\ref{Ent:Prepare}. Then, they measure some of them in the standard basis
 in the signal quality control test in step~\ref{Ent:QCTest}. More importantly,
 since all the LOCC2 EP and PEC privacy amplification procedures in
 step~\ref{Ent:PriAmp} map standard basis to standard basis, we can imagine
 conceptually that the final standard basis measurements of their shared secret
 key were performed right at the beginning of step~\ref{Ent:PriAmp}. In this
 way, any quantum eavesdropping strategy used by Eve is reduced to a classical
 probabilistic cheating strategy \cite{lochauqkdsec}.

 Further recall that in step~\ref{Ent:QCTest}, Alice and Bob do not care about
 the measurement outcome of an individual quantum register; they only care
 about the difference between the measurement outcome of Alice and the
 corresponding outcome of Bob. In other words, they apply the projection
 operators
\begin{equation}
 P_a = \sum_{i\in GF(N)} |i,i+a\rangle \,\langle i,i+a| \label{E:P_form_ind}
\end{equation}
 to the randomly selected quantum registers they share in the set $S_0$. These
 projection operators can be rewritten in a form involving Bell-like states as
 follows. Define $|\Phi_{ab}\rangle$ to be the Bell-like state $\sum_{i\in
 GF(N)} |i\rangle \otimes X_a Z_b |i\rangle / \sqrt{N} \equiv \sum_{i\in GF(N)}
 \omega_p^{\textrm{Tr} (i b)} |i,i+a\rangle / \sqrt{N}$. Then the projection
 operator $P_a$ can also be written as
\begin{equation}
 P_a = \sum_{i\in GF(N)} |\Phi_{ai}\rangle \,\langle \Phi_{ai}| .
 \label{E:P_form_Bell}
\end{equation}
 In a similar way, Alice and Bob apply the projection operators $T^{-i} P_a
 T^i$ to the set $S_i$ for all $i$. Now, it is straight-forward to check that
 the unitary operator $T$ maps Bell-like states to Bell-like states. Combining
 with Eqs.~(\ref{E:P_form_ind}) and~(\ref{E:P_form_Bell}), the signal quality
 control test in step~\ref{Ent:QCTest} of Scheme~A can be regarded as an
 effective random sampling test for the fidelity of the pairs as $|\Phi_{00}
 \rangle \equiv \sum_{i\in GF(N)} |ii\rangle / \sqrt{N}$.

 At this point, classical sampling theory can be used to estimate the quantum
 channel error and hence the eavesdropping rate of the classical probabilistic
 cheating strategy used by Eve as well as the fidelity of the remaining pairs
 as $|\Phi_{00}\rangle$.

\par\medskip
\begin{Lem}[Adapted from Lo, Chau and Ardehali \cite{biasedbb84}]
 Suppose that immediately after step~\ref{Ent:Prepare} in Scheme~A, Alice and
 Bob share $L_i$ pairs of particles in the set $S_i$, namely, those particles
 that are evolved under $T^i$ and then $T^{-i}$. Suppose further that Alice
 and Bob randomly pick $\mbox{O} (\log [ 1/ \epsilon] / \delta^2) \lesssim 0.01
 L_i$ out of the $L_i$ pairs for testing in step~\ref{Ent:QCTest}. Define the
 estimated channel standard basis measurement error rate $\hat{e}_i$ to be the
 portion of tested pairs whose measurement results obtained by Alice and Bob
 differ. Denote the channel standard basis measurement error rate for the set
 $S_i$ by $e_i$.  Then, the probability that $|e_i - \hat{e}_i| > \delta$ is of
 the order of $\epsilon$ for any fixed $\delta > 0$. \label{Lem:Sampling}
\end{Lem}
\begin{proof}
 Using earlier discussions in this subsection, the problem depicted in this
 lemma is equivalent to a classical random sampling problem without
 replacement whose solution follows directly from Lemma~1 in
 Ref.~\cite{biasedbb84}.
\end{proof}

\par\medskip
 Lemma~\ref{Lem:Sampling} assures that by randomly choosing $\mbox{O} (\log [1
 / \epsilon] / \delta^2)$ out of $L_i$ pairs to test, the unbiased estimator
 $\hat{e}_i$ cannot differ from the actual channel standard basis measurement
 error rate $e_i$ significantly. More importantly, the number of particle pairs
 they need to test is independent of $L_i$. Therefore, in the limit of large
 $L_i$ (and hence large $L$), randomly testing a negligibly small portion of
 quantum particle pairs is sufficient for Alice and Bob to estimate with high
 confidence the channel standard basis measurement error rate in the set $S_i$
 \cite{biasedbb84}. In addition, the QER of the remaining untested particle
 pairs is the same as that of $\bigcup_{i=0}^N S_i$ in the large $L$ limit.

\par\medskip
\begin{Thrm}
 Using the notation in Lemma~\ref{Lem:Sampling}, $\sum_{i=0}^N \hat{e}_i / N$
 is a reliable estimator of the upper bound of the QER. Specifically, the
 probability that the QER exceeds $\sum_{i=0}^N \hat{e}_i / N + (N+1) \delta /
 N$ is less than $\epsilon (N+1)$. \label{Thrm:Sampling}
\end{Thrm}
\begin{proof}
 Recall that Eve does not know the choice of unitary operators applied by Alice
 and Bob in step~\ref{Ent:Prepare} in Scheme~A. Hence, in the limit of large
 $L$, the $X_a Z_b$ error rate in the set $S_0$ is equal to that of $T^{-k} X_a
 Z_b T^k$ in the set $S_k$. Therefore, this theorem follows directly from
 Corollary~\ref{Cor:Sequence} and Lemma~\ref{Lem:Sampling}.
\end{proof}

\par\medskip
 To summarize, once the signal quality control test in step~\ref{Ent:QCTest} of
 Scheme~A is passed, Alice and Bob have high confidence (of at least $(1-
 \epsilon)$) that the QER of the remaining untested particle pairs is small.

 Before leaving this subsection, I would like to point out that one can
 estimate the QER in a more aggressive way. Specifically, Alice and Bob do not
 simply know whether the measurement results of each tested pair are equal, in
 fact they know the difference between their measurement results in each tested
 pair. They may exploit this extra piece of information to better estimate the
 probability of $X_a Z_b$ error in the signal for each $a,b\in GF(N)$. Such
 estimation helps them to devise tailor-made privacy amplification schemes that
 tackle the specific kind of error caused by channel noise and Eve. While this
 methodology will be useful in practical QKD, I shall not pursue this direction
 further here as the aim of this paper is the worst-case cryptanalysis in the
 limit of large number of quantum particle transfer $L$.

\subsection{Security Of Privacy Amplification \label{Subsec:PA}}
\par\medskip
\begin{Def}
 We denote the $X_a Z_b$ error rate of the quantum particles shared by Alice
 and Bob just before step~\ref{Ent:PriAmp} in Scheme~A by $e_{a,b}$. And when
 there is no possible confusion in the subscript, we shall write $e_{ab}$
 instead of $e_{a,b}$. Similarly, we denote the $X_a Z_b$ error rate of the
 resultant quantum particles shared by them after $k$ rounds of LOCC2 EP by
 $e_{a,b}^{k\,\textrm{EP}}$ or $e_{ab}^{k\,\textrm{EP}}$. Suppose further that
 Alice and Bob perform PEC using the $[r,1,r]_N$ majority vote code after $k$
 rounds of LOCC2 EP. We denote the resultant $X_a Z_b$ error rate by
 $e_{a,b}^\textrm{PEC}$ or $e_{ab}^\textrm{PEC}$. \label{Def:e_ab}
\end{Def}

\par\medskip
 Recall that Alice and Bob randomly and independently apply $T^i$ and $T^{-j}$
 to each transmitted quantum register. More importantly, their choices are
 unknown to Eve when the quantum particle is traveling in the insecure channel.
 Let ${\mathcal E}$ be the quantum operation that Eve applies to the quantum
 particles in the set $\bigcup_{i=0}^N S_i$. (In other words, ${\mathcal E}$ is
 a completely positive convex-linear map acting on the set of density matrices
 describing the quantum particle pairs to which Alice and Bob has applied $T^j$
 and $T^{-j}$ respectively for some $j$. Moreover, the trace of ${\mathcal E}$
 is between 0 and 1.) After Alice and Bob have publicly announced their choices
 of quantum operations, the quantum particle pairs in $\bigcup_{i=0}^N S_i$ had
 equal chance of suffering from $(\otimes_j T^{-i_j}) {\mathcal E} (\otimes_j
 T^{i_j})$ where $0\leq i_j \leq N$. Note that the index $j$ in the tensor
 product in the above expression runs over all particles pairs in
 $\bigcup_{i=0}^N S_i$. Besides, the privacy amplification procedure in
 step~\ref{Ent:PriAmp} is performed irrespective to which set $S_i$ the
 particle belong to. Therefore, the QER satisfies the constraints
\begin{equation}
 \sum_{i,j\in GF(N)} e_{ij} = 1 \label{E:Constraint_e_ab_trivial}
\end{equation}
 and
\begin{equation}
 e_{ab} = e_{a'b'} \mbox{~if~} (a,b) \sim (a',b') . \label{E:Constraint_e_ab}
\end{equation}

 After knowing the initial conditions for the QER, I am going to investigate
 the effect of LOCC2 EP on the QER.

\par\medskip
\begin{Lem}
 In the limit of a large number of transmitted quantum registers, $e_{ab}^{k
 \,\textrm{EP}}$ is given by
 \begin{equation}
  e_{ab}^{k \,\textrm{EP}} = \frac{\sum_{c_0,\ldots ,c_{2^k-2}} e_{ac_0}
  e_{ac_1} \cdots e_{ac_{2^k-2}} e_{a,b-c_0-c_1-\cdots -c_{2^k-2}}}{\sum_{i\in
  GF(N)} \left( \sum_{j\in GF(N)} e_{ij} \right)^{2^k}} ~. \label{E:iter_e_ab}
 \end{equation}
 Moreover, in this limit, $e_{ab}^{k \,\textrm{EP}} = e_{-a,-b}^{k
 \,\textrm{EP}}$ for all $a,b\in GF(N)$ and $k\in {\mathbb N}$. \label{Lem:EP}
\end{Lem}
\begin{proof}
 Suppose the control and target registers in Bob's laboratory suffer from $X_a
 Z_b$ and $X_{a'} Z_{b'}$ errors respectively. (In contrast, those in Alice's
 hand are error-free as they never pass through the insecure noisy channel.)
 Then after applying the unitary operation in Eq.~(\ref{E:GenCNOT}), the errors
 in the control and target registers will become $X_a Z_{b+b'}$ and $X_{a'-a}
 Z_{b'}$ respectively.

 In the limit of large number of transmitted quantum registers, the covariance
 between probabilities of picking any two distinct quantum register tends to
 zero. Besides, the covariance between probabilities of picking any two
 distinct pairs of quantum registers also tends to zero. Hence, in this limit,
 the expectation value of the $X_a Z_b$ error rate just after applying the
 unitary operation in Eq.~(\ref{E:GenCNOT}) can be computed by assuming that
 the error in every control and target register pair is independent. Moreover,
 the variance of the $X_a Z_b$ error rate tends to zero in this limit.

 To show that Eq.~(\ref{E:iter_e_ab}) is valid, let us recall that Alice and
 Bob keep their control registers only when the measurement results of their
 corresponding target registers agree. In other words, they keep the control
 registers only when $a = a'$. Thus, once the control register in Bob's
 laboratory is kept, it will suffer an error $X_d Z_c$ where $d = a$ and $c = b
 + b'$. Therefore, in the limit of a large number of transmitted quantum
 registers, the number of quantum registers remains after $(k+1)$ rounds of
 LOCC2 EP is proportional to $\sum_{i\in GF(N)} ( \sum_{j\in GF(N)} e_{ij}^{k
 \,\textrm{EP}} )^2$. Similarly, the number of quantum registers suffering from
 $X_a Z_b$ error after $(k+1)$ rounds of LOCC2 EP is proportional to $\sum_{c
 \in GF(N)} e_{ac}^{k \,\textrm{EP}} e_{a,b-c}^{k \,\textrm{EP}}$. More
 importantly, the two proportionality constants are the same. Therefore,
 \begin{equation}
  e_{ab}^{(k+1) \,\textrm{EP}} = \frac{\sum_{c \in GF(N)} e_{ac}^{k
  \,\textrm{EP}} e_{a,b-c}^{k \,\textrm{EP}}}{\sum_{i\in GF(N)} \left( \sum_{j
  \in GF(N)} e_{ij}^{k \,\textrm{EP}} \right)^2} \label{E:e1_ab}
 \end{equation}
 for all $k\in {\mathbb N}$. Eq.~(\ref{E:iter_e_ab}) can then be proven by
 mathematical induction on $k$. (It is easier to use mathematical induction to
 prove the validity of the numerator in Eq.~(\ref{E:iter_e_ab}) and then use
 Eq.~(\ref{E:Constraint_e_ab_trivial}) to determine the denominator.)

 To show that $e_{ab}^{k \,\textrm{EP}} = e_{-a,-b}^{k \,\textrm{EP}}$, I only
 consider the case of $p>2$ since the assertion is trivially true when $p=2$.
 From Corollary~\ref{Cor:EquivalentClass} and Eq.~(\ref{E:Constraint_e_ab}), we
 have $e_{ab} = e_{-a,-b}$. Inductively, assuming the validity of the assertion
 for $k$, then $e_{ab}^{(k+1) \,\textrm{EP}} = \sum_{c\in GN(N)} e_{ac}^{k
 \,\textrm{EP}} e_{a,b-c}^{k \,\textrm{EP}} / D_k = \sum_{c\in GF(N)}
 e_{-a,-c}^{k \,\textrm{EP}} e_{-a,-b+c}^{k \,\textrm{EP}} / D_k =
 e_{-a,-b}^{(k+1)\,\textrm{EP}}$, where $D_k = \sum_{i\in GF(N)} ( \sum_{j\in
 GF(N)} e_{ij}^{k\,\textrm{EP}} )^2$. Hence, the lemma is proved.
\end{proof}

\par\medskip
 Eq.~(\ref{E:iter_e_ab}) in Lemma~\ref{Lem:EP} can be expressed in a more
 compact and useful form below.

\par\medskip
\begin{Cor}
 Any element $a\in GF(N) \equiv GF(p^n)$ can be expressed as a degree $(n-1)$
 polynomial $a_0 + a_1 x + \cdots + a_{n-1} x^{n-1}$ in $GF(p)[x]$. With this
 notation in mind, $e_{ab}^{k \,\textrm{EP}}$ in Eq.~(\ref{E:iter_e_ab}) can be
 rewritten as 
 \begin{eqnarray}
  e_{ab}^{k \,\textrm{EP}} & = & \left\{ \sum_{m_0,\ldots ,m_{n-1}=0}^{p-1}
   \cos \left( \frac{2 \pi \sum_{i=0}^{n-1} m_i b_i}{p} \right) \times \right.
   \nonumber \\
  & & \left. \left[ \sum_{j\in GF(N)} e_{aj} \cos \left( \frac{2\pi \sum_{i=
   0}^{n-1} m_i j_i}{p} \right) \right]^{2^k} \right\} \times \nonumber \\
  & & \left[ N \sum_{i\in GF(N)} \left( \sum_{j\in GF(N)} e_{ij} \right)^{2^k}
   \right]^{-1} . \label{E:e_ab}
 \end{eqnarray}
 In particular, if $e_{ab}$ satisfies
 \begin{equation}
  e_{ab} = \left\{ \begin{array}{cl}
   \displaystyle \frac{1-e_{00}}{N+1} & \mbox{if~} (a,b) \sim (0,1) , \\
   \\
   0 & \mbox{if~} (a,b)\not\sim (0,0) \mbox{~and~} (0,1) ,
  \end{array} \right. \label{E:e_ab_min_def}
 \end{equation}
 then for $p=2$,
 \begin{equation}
  e_{00}^{k \,\textrm{EP}} = \frac{(e_{00}+e_{01})^{2^k} + (e_{00}-
  e_{01})^{2^k}}{2[(e_{00}+e_{01})^{2^k} + \sum_{i\neq 0} ( \sum_{j\in GF(N)}
  e_{ij})^{2^k}]} , \label{E:worst_case_e_00_p2}
 \end{equation}
 \begin{equation}
  e_{01}^{k \,\textrm{EP}} = \frac{(e_{00}+e_{01})^{2^k} - (e_{00}-
  e_{01})^{2^k}}{2[(e_{00}+e_{01})^{2^k} + \sum_{i\neq 0} ( \sum_{j\in GF(N)}
  e_{ij})^{2^k}]} \label{E:worst_case_e_01_p2}
 \end{equation}
 and
 \begin{equation}
  e_{0b}^{k \,\textrm{EP}} = 0 \mbox{~for~} b\neq 0,1 .
  \label{E:worst_case_e_0b_p2}
 \end{equation}
 \label{Cor:EP}
\end{Cor}
\begin{proof}
 The numerator of Eq.~(\ref{E:iter_e_ab}) is equal to the sum of coefficients
 of the terms in the form $x_0^{m_0} x_1^{m_1} \cdots x_{n-1}^{m_{n-1}}$ in the
 polynomial $(\sum_{j\in GF(N)} e_{aj} x_0^{j_0} x_1^{j_1} \cdots
 x_{n-1}^{j_{n-1}})^{2^k}$ where $m_i = -b_i \bmod p$ for all $i$. This sum is
 in turn equal to $\sum_{x_0,\ldots ,x_{m-1} = 1,\omega_p,\ldots
 ,\omega_p^{p-1}}$ $x_0^{-b_0} \cdots x_{n-1}^{-b_{n-1}}$ $(\sum_{j\in GF(N)}
 e_{aj} x_0^{j_0} \cdots x_{n-1}^{j_{n-1}})^{2^k} / N$. Since the imaginary
 part of the above sum is zero, I arrive at the expression in
 Eq.~(\ref{E:e_ab}).

 The proof of the remaining parts of this lemma now follow directly from
 Eq.~(\ref{E:e_ab}) and Corollary~\ref{Cor:EquivalentClass}.
\end{proof}

\par\medskip
 Lemma~\ref{Lem:EP} and Corollary~\ref{Cor:EP} generalize a similar result for
 qubits \cite{qkd2waylocc,sixstateexact}. In fact, the effect of LOCC2 EP is to
 reduce errors in the form $X_a Z_b$ with $a\neq 0$ at the expense of possibly
 increasing errors in the form $Z_c$ with $c\neq 0$. I further remark that in
 case $L$ is finite, $e_{ab}^{k \,\textrm{EP}}$ is determined by solving the
 classical problem of randomly pairing $N^2$ kinds of balls in an urn
 containing $2r\ell$ balls. Therefore, $e_{ab}^{k \,\textrm{EP}}$ is related to
 the so-called multivariate hypergeometric distribution whose theory is
 reviewed extensively in Ref.~\cite{multidist}.

\par\medskip
 In the qubit case, that is when $N = p = 2$,
 Eqs.~(\ref{E:Constraint_e_ab_trivial}) and~(\ref{E:Constraint_e_ab}) demand
 that $e_{01} = e_{10} = e_{11} = (1-e_{00})/3$. In other words, the evolution
 of QER under the action of LOCC2 EP depends on a single parameter, namely,
 $e_{00}$. Nevertheless, the situation is more complicated when $N > 2$ because
 $e_{ab}^{k \,\textrm{EP}}$ depends on more than one parameter. Fortunately, as
 we shall see later on, it is possible to determine the worst case scenario for
 $e_{ab}$ when the number of rounds of LOCC2 EP, $k$, is sufficiently large
 when $p=2$.
 
\par\medskip
\begin{Lem}
 The following two statements hold provided that either (1)~$p = 2$ and $e_{00}
 > 1/(N+2)$ or (2)~$p > 2$ and $e_{00} > 2/(N+3)$.

 (a) The maximum term in the denominator of Eq.~(\ref{E:e_ab}) is $(\sum_{j\in
 GF(N)} e_{0j})^{2^k}$.

 (b) $e_{00}^{k \,\textrm{EP}} > e_{0b}^{k \,\textrm{EP}}$ whenever $b\neq 0$.
 \label{Lem:Dominance}
\end{Lem}
\begin{proof}
 To prove the first statement, I first consider the $p = 2$ case. Using
 Corollary~\ref{Cor:EquivalentClass} plus the two constraints in
 Eqs.~(\ref{E:Constraint_e_ab_trivial}) and~(\ref{E:Constraint_e_ab}), we have
 $e_{00} > (1-e_{00})/(N+1) = \sum_{j\neq 0} e_{0j} \geq e_{ab}$ for all $(a,b)
 \neq (0,0)$. Hence, Corollary~\ref{Cor:EquivalentClass} demands that $\sum_j
 (e_{0j} - e_{ij}) \geq e_{00} - \sum_{j\neq 0} e_{0j} > 0$ for all $i\neq 0$.
 By the same argument, in the $p > 2$ case, $\sum_j (e_{0j}-e_{ij}) \geq
 e_{00} - 2(1-e_{00})/(N+1) > 0$ for all $i\neq 0$.

 To prove the second statement, I express $e_{00}^{k \,\textrm{EP}} - e_{0b}^{k
 \,\textrm{EP}}$ in terms of $e_{ij}^{(k-1) \,\textrm{EP}}$ by invoking
 Eq.~(\ref{E:e1_ab}). The denominator of this expression is positive and the
 numerator is given by
\begin{eqnarray}
 & & \sum_{c\in GF(N)} e_{0c}^{(k-1) \,\textrm{EP}} \left[ e_{0,-c}^{(k-1)
  \,\textrm{EP}} - e_{0,b-c}^{(k-1) \,\textrm{EP}} \right] \nonumber \\
 & = & \sum_{c\in GF(N)} e_{0c}^{(k-1) \,\textrm{EP}} \left[ e_{0c}^{(k-1)
  \,\textrm{EP}} - e_{0,b-c}^{(k-1) \,\textrm{EP}} \right] \nonumber \\
 & = & \frac{1}{2} \sum_{c\in GF(N)} \left[ e_{0c}^{(k-1) \,\textrm{EP}} -
  e_{0,b-c}^{(k-1) \,\textrm{EP}} \right]^2 , \label{E:e_00_minus_e_0b}
\end{eqnarray}
 where I have used Lemma~\ref{Lem:EP} to arrive at the second line. Therefore,
 $e_{00}^{k \,\textrm{EP}} \geq e_{0b}^{k \,\textrm{EP}}$ for all $b$. In fact,
 our assumption on the value of $e_{00}$ implies $e_{00} > e_{0b}$ for all $b
 \neq 0$. Hence from Eq.~(\ref{E:e_00_minus_e_0b}), statement~(b) holds for
 $k=1$. The validity of statement~(b) for all $k\in {\mathbb Z}^+$ can then be
 shown by mathematical induction on $k$.
\end{proof}

\par\medskip
\begin{Thrm}
 In the limit of large number of quantum particle transmitted from Alice to
 Bob, the $X_a Z_b$ error rate after PEC $e_{ab}^\textrm{PEC}$ using
 $[r,1,r]_N$ majority vote code satisfies
 \begin{equation}
  \sum_{i\neq 0} \sum_{j\in GF(N)} e_{ij}^\textrm{PEC} \leq r \sum_{i\neq 0}
  \sum_{j\in GF(N)} e_{ij}^{k \,\textrm{EP}} . \label{E:PEC_spin_flip}
 \end{equation}
 Moreover, if $p=2$ and $e_{00} > 1/(N+2)$, then
 \begin{eqnarray}
  & & \sum_{i\in GF(N)} \sum_{j\neq 0} e_{ij}^\textrm{PEC} \nonumber \\
  & \leq & (N-1) \left[ 1 - \frac{(e_{00}- \frac{1-e_{00}}{N+1})^{2^{k+1}}}{2
  (e_{00}+\frac{1-e_{00}}{N+1})^{2^{k+1}}} \right]^r
  \label{E:PEC_inequality_p2}
 \end{eqnarray} 
 as $k\rightarrow\infty$. \label{Thrm:PEC}
\end{Thrm}
\begin{proof}
 Recall that the error syndrome of the $[r,1,r]_N$ majority vote code is
 \begin{equation}
  \left[ \begin{array}{ccccc}
   1 & -1 \\
   1 & & -1 \\
   \vdots & & & \ddots \\
   1 & & & & -1
  \end{array} \right] . \label{E:majority_vote_syndrome}
 \end{equation}
 Therefore, after measuring the (phase) error syndrome, $Z_b$ error stays on
 the control register while $X_a$ error propagates from the control as well as
 all target registers to the resultant control quantum register \cite{hd_ftqc}.
 Specifically, suppose the error on the $i$th quantum register is $X_{a_i}
 Z_{b_i}$ for $i=1,2,\ldots ,r$. Then, after measuring the error syndrome, the
 resultant error in the remaining control register equals $X_{a_1 + \cdots +
 a_r} Z_{b_1}$. Consequently, upon PEC, the error in the remaining register is
 $X_{a_1 + \cdots + a_r} Z_b$ where $b$ is the majority of $b_i$ ($i=1,2,\ldots
 ,r$). In other words, after PEC, spin flip error rates are increased by at
 most $r$ times. Hence, Eq.~(\ref{E:PEC_spin_flip}) holds.

 By the same argument used in Lemma~\ref{Lem:EP}, in the limit of large number
 of quantum register transfer, the rate of any kind of phase error after PEC,
 $\sum_{i\in GF(N)} \sum_{j\neq 0} e_{ij}^\textrm{PEC}$, satisfies
 \begin{eqnarray}
  & & \sum_{i\in GF(N)} \sum_{j\neq 0} e_{ij}^\textrm{PEC} \nonumber \\
  & \leq & (N-1) \max \{ \textrm{Pr} \,( \textrm{the number of registers
   suffering} \nonumber \\
  & & ~\textrm{from error in the form } X_a Z_1 \textrm{ is greater than or}
   \nonumber \\
  & & ~\textrm{equal to those suffering from error in the form } X_a
   \nonumber \\
  & & ~\textrm{when drawn from a random sample of } r \textrm{ registers,}
   \nonumber \\
  & & ~\textrm{given a fixed } e_{00}) \} , \label{E:PEC_inequality1}
 \end{eqnarray}
 where the maximum is taken over all possible probabilities with different
 $e_{ab}$'s satisfying the constraints in
 Eqs.~(\ref{E:Constraint_e_ab_trivial}) and~(\ref{E:Constraint_e_ab}). I denote
 the sum $\sum_{i\in GF(N)} e_{ib}^{k\,\textrm{EP}}$ by $e_{Z_b}^{k
 \,\textrm{EP}}$. Then,
 \begin{eqnarray}
  & & \sum_{i\in GF(N)} \sum_{j\neq 0} e_{ij}^\textrm{PEC} \nonumber \\
  & \leq & (N-1) \max \{ \sum_{s=0}^r \binom{r}{s} (1-e_{Z_0}^{k \,\textrm{EP}}
   - e_{Z_1}^{k \,\textrm{EP}})^{r-s} \times \nonumber \\
  & & ~(e_{Z_0}^{k \,\textrm{EP}} + e_{Z_1}^{k \,\textrm{EP}})^s \,\textrm{Pr}
   ( \textrm{the number of registers suffering} \nonumber \\
  & & ~\textrm{from error in the form } X_a Z_1 \textrm{ is greater than or}
   \nonumber \\
  & & ~\textrm{equals to those suffering from error in the from } X_a \nonumber
   \\
  & & ~\textrm{when drawn from a random sample of } s \textrm{ registers,}
   \nonumber \\
  & & ~\textrm{given that these } s \textrm{ registers are suffering from}
   \textrm{ error} \nonumber \\
  & & ~\textrm{in the form } X_a Z_b \textrm{ for } b = 0,1 \textrm{ and given}
   \textrm{ a fixed } e_{00}) \} \nonumber \\
  & \leq & (N-1) \max \{ \sum_{s=0}^r \binom{r}{s} (1-e_{Z_0}^{k \,\textrm{EP}}
   -e_{Z_1}^{k \,\textrm{EP}})^{r-s} \times \nonumber \\
  & & ~(e_{Z_0}^{k \,\textrm{EP}} + e_{Z_1}^{k \,\textrm{EP}})^s \exp \left[ -2
   s \left( \frac{1}{2}-\frac{e_{Z_1}^{k \,\textrm{EP}}}{e_{Z_0}^{k
   \,\textrm{EP}} + e_{Z_1}^{k \,\textrm{EP}}} \right)^2 \right] \} \nonumber
   \\
  & = & (N-1) \max \{ \left\{ 1 - (e_{Z_0}^{k \,\textrm{EP}} + e_{Z_1}^{k
   \,\textrm{EP}}) \times \right. \nonumber \\
  & & ~\left. \left[ e^{-2 [ 1/2 - e_{Z_1}^{k \,\textrm{EP}}/(e_{Z_0}^{k
   \,\textrm{EP}}+e_{Z_1}^{k \,\textrm{EP}})]^2} - 1 \right] \right\}^r \}
   \nonumber \\
  & \leq & (N-1) \max \{ \left[ 1- 2 t (e_{Z_0}^{k \,\textrm{EP}} + e_{Z_1}^{k
   \,\textrm{EP}}) \times \right. \nonumber \\
  & & ~\left. \left( \frac{1}{2} - \frac{e_{Z_1}^{k \,\textrm{EP}}}{e_{Z_0}^{k
   \,\textrm{EP}}+e_{Z_1}^{k \,\textrm{EP}}} \right)^2 \right]^r \}
   \label{E:PEC_inequality2}
 \end{eqnarray}
 where $t\rightarrow 1$ as $k\rightarrow\infty$. Note that I have used
 Eq.~(1.2.5) in Ref.~\cite{Roman} to arrive at the second inequality above.
 (Eq.~(1.2.5) is applicable because Lemma~\ref{Lem:Dominance} implies that
 $e_{Z_0}^{k \,\textrm{EP}} > e_{Z_1}^{k \,\textrm{EP}}$ for a sufficiently
 large $k$.)

 Since $e_{00}$ satisfies $p=2$ and $e_{00} > 1/(N+2)$,
 Lemma~\ref{Lem:Dominance} tells us that $(\sum_{j\in GF(N)} e_{0j})^{2^k}$ is
 the dominant term in the denominator of Eq.~(\ref{E:e_ab}) when $k$ is
 sufficiently large. Thus, using Eq.~(\ref{E:e_ab}), it is easy to check that
 both $e_{Z_1}^{k\,\textrm{EP}} / e_{Z_0}^{k\,\textrm{EP}}$ and $e_{Z_0}^{k
 \,\textrm{EP}} + e_{Z_1}^{k \,\textrm{EP}}$ are maximized if $e_{ab} = (1-
 e_{00})/(N+1)$ for all $(a,b) \sim (0,1)$ when subjected to the following two
 constraints: (1)~$e_{00}$ is fixed; and
 (2)~Eqs.~(\ref{E:Constraint_e_ab_trivial}) and~(\ref{E:Constraint_e_ab}) are
 satisfied. Therefore, the last line of Eq.~(\ref{E:PEC_inequality2}) is
 maximized if Eq.~(\ref{E:e_ab_min_def}) holds. Consequently,
 Eqs.~(\ref{E:worst_case_e_00_p2}) and~(\ref{E:worst_case_e_0b_p2}) imply the
 validity of Eq.~(\ref{E:PEC_inequality_p2}).
\end{proof}

\par\medskip
 The above theorem tells us that the effect of PEC is reducing errors in the
 form $X_a Z_b$ with $b\neq 0$ at the expense of possibly increasing errors in
 the form $X_c$ with $c\neq 0$. For this reason, powerful signal privacy
 amplification procedure can be constructed by suitably combining LOCC2 EP and
 PEC.

\par\medskip
 Now, I am going to prove the unconditional security of Scheme~A.

\par\medskip
\begin{Thrm}
 Let $N = p^n$ be a prime power, $\epsilon_p$, $\epsilon_I$ and $\delta$ be
 three arbitrarily small but fixed positive numbers. Define
 \begin{equation}
  e^\textrm{QER} = \frac{(N+1)(\sqrt{5}-2)}{1+(N+1)(\sqrt{5}-2)} \mbox{~~for~}
  p=2 . \label{E:critical_p2}
 \end{equation}
 Then, the entanglement-based QKD Scheme~A involving the transfer of
 $N$-dimensional quantum particles is unconditionally secure with security
 parameters $(\epsilon_p,\epsilon_I)$ when the number of quantum register
 transfer $L \equiv L(\epsilon_p,\epsilon_I,\delta)$ is sufficiently large.
 Specifically, provided that Alice and Bob abort the scheme whenever the
 estimated QER in step~\ref{Ent:QCTest} is greater than $(e^\textrm{QER} -
 \delta)$, then the secret key generated by Alice and Bob is provably secure in
 the $L\rightarrow\infty$ limit. In fact, if Eve uses an eavesdropping strategy
 with at least $\epsilon_p$ chance of passing the signal quality test stage in
 step~\ref{Ent:QCTest}, the mutual information between Eve's measurement
 results after eavesdropping and the final secret key is less than
 $\epsilon_I$. In this respect, Scheme~A tolerates asymptotically up to
 $e^\textrm{QER}$ QER. \label{Thrm:Uncond_Sec_A}
\end{Thrm}
\begin{proof}
 Since $L \gg (N+1)^4$ $\log [(N+1)/\epsilon_p]/\delta^2 N^2$, therefore by
 applying Lemma~\ref{Lem:Sampling} and Theorem~\ref{Thrm:Sampling}, I conclude
 that by testing $\mbox{O} ([N+1]^2 \log \{ [N+1]/\epsilon_p \} / \delta^2
 N^2)$ pairs, any eavesdropping strategy that causes a QER higher than
 $e^\textrm{QER}$ has less than $\epsilon_p$ chance of passing the signal
 quality test stage in step~\ref{Ent:QCTest} of Scheme~A. (Similarly, if the
 QER is less than $(e^\text{QER}-2\delta)$, it has at least $(1-\epsilon_p)$
 chance of passing step~\ref{Ent:QCTest}. As $\delta$ can be chosen to be
 arbitrarily small, the signal quality test stage in step~\ref{Ent:QCTest} of
 Scheme~A is not overly conservative.)

 Now, suppose that Alice and Bob arrive at the signal privacy amplification
 stage in step~\ref{Ent:PriAmp} of Scheme~A. Since $L\rightarrow\infty$, the
 quantum particle pairs used in the signal quality test stage in
 step~\ref{Ent:QCTest} do not affect the error rates $e_{ab}$'s of the
 remaining untested particle pairs.

 First, I consider the case when $p=2$. After applying $k$ rounds of LOCC2 EP,
 Alice and Bob may consider picking $r$ used in the majority vote PEC to be
 $\epsilon_I / 2\sum_{i\in GF(N)} \sum_{j\neq 0} e_{ij}^{k \,\textrm{EP}}$. In
 the limit of $k\rightarrow \infty$, Corollaries~\ref{Cor:EquivalentClass}
 and~\ref{Cor:EP} imply that in the worst case scenario, there are at most two
 distinct $b=b(a)$ and $b'= b'(a)$ such that $e_{ab}, e_{ab'} > 0$ for all $a
 \neq 0$. Hence, $r$ can be chosen to be
 \begin{equation}
  r \approx \frac{\epsilon_I [e_{00}+ (1-e_{00})/(N+1)]^{2^k}}{\ell N [2(1-
  e_{00})/(N+1)]^{2^k}} \label{E:r_bound_p2}
 \end{equation}
 whenever $e_{00} > 1/(N+2)$, where $\ell$ is the number of quantum particle
 pairs Alice and Bob share immediately after the PEC procedure in
 step~\ref{Ent:PriAmp}b. Besides, $r\rightarrow\infty$ in the $k\rightarrow
 \infty$ limit. So, from Eqs.~(\ref{E:PEC_spin_flip})
 and~(\ref{E:PEC_inequality_p2}) in Theorem~\ref{Thrm:PEC}, the QER of the
 remaining quantum registers after PEC, $e^\textrm{final}$ is upper-bounded by
 \begin{equation}
  e^\textrm{final} < \frac{\epsilon_I}{2\ell} + (N-1) \exp \left[
  \frac{-\epsilon_I (e_{00} - \frac{1-e_{00}}{N+1})^{2^{k+1}}}{2\ell N (e_{00}
  +\frac{1-e_{00}}{N+1})^{2^k} [\frac{2(1-e_{00})}{N+1} ]^{2^k}} \right] .
  \label{E:QER_p2}
 \end{equation} 
 In other words, $e^\textrm{final} < \epsilon_I / \ell$ provided that
 \begin{equation}
  \left[ e_{00} - \frac{1-e_{00}}{N+1} \right]^2 > \frac{2(1-e_{00})}{N+1}
  \,\left[ e_{00} + \frac{1-e_{00}}{N+1} \right] . \label{E:QER_Condition_p2_1}
 \end{equation}
 This condition is satisfied if and only if 
 \begin{equation}
  e_{00} > \frac{1}{1+(N+1)(\sqrt{5}-2)} . \label{E:QER_Condition_p2_2}
 \end{equation}
 It is easy to verify that the constraint in Eq.~(\ref{E:QER_Condition_p2_2})
 is consistent with the assumption that $e_{00} > 1/(N+2)$. Hence, provided
 that the initial QER satisfies
 \begin{equation}
  \sum_{(i,j)\neq (0,0)} e_{ij} < \frac{(N+1)(\sqrt{5}-2)}{1+(N+1)(\sqrt{5}-2)}
   = e^\textrm{QER} , \label{E:QER_Condition_p2_3}
 \end{equation}
 the fidelity of the $\ell$ quantum particle pairs shared between Alice and Bob
 immediately before they perform standard basis measurement to obtain their
 secret key is at least $1-e^\textrm{final} > 1-\epsilon_I/\ell$. By
 Footnote~28 in \cite{lochauqkdsec}, the mutual information between Eve's final
 measurement result after eavesdropping and the final secret key is at most
 $\epsilon_I$. Thus, if Alice and Bob abort the scheme if the estimated QER in
 step~\ref{Ent:QCTest} exceeds $(e^\textrm{QER}-\delta)$, the secret key
 generated is provably secure. More importantly, the scheme is unconditionally
 secure with security parameters $(\epsilon_p,\epsilon_I)$.
\end{proof}

\par\medskip
 A few remarks are in order. First, the unconditional security of Scheme~A for
 $p>2$ can be proven in a similar way. However, the computation of
 $e^\textrm{QER}$ is getting messy as the condition for minimizing
 $e^\textrm{QER}$ turns out to be $N$ dependent.
 
 Second, from Corollary~\ref{Cor:Sequence}, when $p=
 2$, $GF(N)/\sim = \{ (0,b)/ \sim : b\in GF(N) \}$ and hence the ratio between
 QER and SBMER for any kind of eavesdropping attacks equals $(N+1) : N$. In
 contrast, when $p>2$, such a ratio varies between $(N+1) : (N-1)$ and $1 : 1$.
 Combining these observations with Theorem~\ref{Thrm:Uncond_Sec_A}, I conclude
 that the maximum tolerable SBMER for Scheme~A is given by
\begin{equation}
 e^\textrm{SBMER} = \left\{ \begin{array}{cl}
  \frac{N e^\textrm{QER}}{N+1} & \mbox{if~} p=2, \\ \\
  \frac{(N-1) e^\textrm{QER}}{N+1} & \mbox{if~} p>2 .
 \end{array} \right. \label{E:SBMER_QER}
\end{equation}
 In addition, if $p=2$, Corollary~\ref{Cor:EquivalentClass} implies that there
 is a unique $a\neq 0$ such that $(0,1) \sim (a,b) \sim (a,b')$ for some $b\neq
 b'$. Hence, no matter which bijective map Alice and Bob use to convert their
 standard basis measurement result of an $N$-dimensional quantum particle into
 a $\log_2 N$-bit string, the ratio between QER and BER is at least $(N+1) : (1
 + 0.5 N \log_2 N) / \log_2 N$. Consequently, the maximum tolerable BER for
 Scheme~A is given by
\begin{equation}
 e^\textrm{BER} = e^\textrm{SBMER} \left( \frac{1}{2} + \frac{1}{N\log_2 N}
 \right) . \label{E:BER_QER}
\end{equation}
 I tabulate the tolerable SBMER and BER in Table~\ref{T:SBMER_BER}. However, I
 must emphasize once again that according to the discussions in
 Subsection~\ref{Subsec:Error_Rates}, we \emph{should not} and \emph{cannot}
 deduce the relative error tolerance capability from Table~\ref{T:SBMER_BER}.
 
\begin{table}
\begin{center}
\begin{tabular}{|r|c|c|} 
 \hline
 N & Tolerable SBMER & Tolerable BER \\ \hline
 2 & 27.64\% & 27.64\% \\
 4 & 43.31\% & 27.07\% \\
 8 & 60.44\% & 32.74\% \\
 16 & 75.34\% & 38.85\% \\
 \hline
\end{tabular}
\end{center}
\caption{The tolerable SBMER and BER for Scheme~A and hence also Scheme~B for
 $2^n\leq 16$. As pointed out in the text, the values of SBMER and BER should
 not be compared directly.}
\label{T:SBMER_BER}
\end{table}
 
 Third, I study the tolerable error rate of Scheme~A as a function of $N$.
 Table~\ref{T:SBMER_BER} shows that the maximum tolerable BER $e^\textrm{BER}$
 for $N=2$ is the same as the one obtained earlier by Chau in
 Ref.~\cite{sixstateexact}. More importantly, $e^\textrm{SBMER}$ increases as
 $n$ increases.

\par\medskip
 Actually, according to Eqs.~(\ref{E:critical_p2}) and
 Eqs.~(\ref{E:SBMER_QER})--(\ref{E:BER_QER}), the tolerable SBMER and BER
 tend to 100\% and 50\% respectively as $2^n\rightarrow\infty$. More precisely,
 as $n\rightarrow\infty$, the tolerable BER for Scheme~A using $2^n$-level
 quantum particles scales as $\approx 1/2 - (3+\sqrt{5})/2^{n+1}$.

 On the other hand, the lemma below set the upper limit for the tolerable SBMER
 for Scheme~A.

\par\medskip
\begin{Lem}
 The tolerable SBMER for Scheme~A is upper-bounded by $(N-1)/(N+1)$ if $p=2$
 and $(N-1)^2/[N(N+1)]$ if $p>2$. In fact, these bounds are set by the
 following interpret-and-resend strategy: Eve randomly and independently
 measures each $N$-dimensional particle in the insecure quantum channel in the
 standard basis $\{ |0\rangle, |1\rangle, \ldots ,|N-1\rangle \}$. Then, she
 records the measurement result and resends the measured particle to Bob.
 \label{Lem:UpperBound_QER_SBMER}
\end{Lem}
\begin{proof}
 The proof follows the idea reported in Ref.~\cite{qkd2waylocc}. Clearly, using
 this intercept-and-resend strategy, no quantum correlation between Alice and
 Bob can survive and hence no provably secure key can be distributed. Thus,
 this eavesdropping strategy set the upper bound for the tolerable SMBER and
 BER for Scheme~A. It is easy to check that the bases $\{ T^i|0\rangle, T^i
 |1\rangle, \ldots ,T^i|N-1\rangle \}$ where $i=0,1,\ldots ,N$ if $p=2$ and $i=
 0,1,\ldots ,(N-1)/2$ if $p>2$ are mutually unbiased. (A proof can be found in
 Lemma~\ref{Lem:MUB} in Section~\ref{Sec:ShorPre} below.) Consequently, if it
 turns out that the measured qubit is prepared in the standard basis, that
 qubit will be accepted by Scheme~A as error-free. In contrast, if the measured
 qubit is not prepared in the standard basis, it has $(N-1)/N$ chance of being
 detected as erroneous. Therefore, the tolerable SBMER is upper-bounded by
 $N/(N+1) \times (N-1)/N = (N-1)/(N+1)$ if $p=2$ and $[(N+1)/2-1]/[(N+1)/2]
 \times (N-1) /N = (N-1)^2/[N(N+1)]$ if $p>2$.
\end{proof}

\par\medskip
 Thus, the difference between the tolerable SBMER and its theoretical upper
 bound tends to zero in the limit of large $N$. So in the limit, the error
 tolerance capability of Scheme~A approaches its maximally allowable value.

 Fourth, readers may wonder why Scheme~A is highly error-tolerant especially
 when $N$ is large. Recall that Eve does not know which particles are in set
 $S_i$ when the particles are transmitted from Alice to Bob. Hence, in the
 limit of large number of quantum particle transfer $L$, $e_{ab}$ satisfies the
 constraints in Eqs.~(\ref{E:Constraint_e_ab_trivial})
 and~(\ref{E:Constraint_e_ab}). This greatly limits the relative occurrence
 rates between different types of quantum errors. At this point, the LOCC2 EP
 becomes a powerful tool to reduce the spin errors at the expense of increasing
 phase errors. Furthermore, provided that the condition in
 Lemma~\ref{Lem:Dominance} holds, $e_{Z_0}^{k \,\textrm{EP}} > e_{Z_b}^{k
 \,\textrm{EP}}$ for all $b\neq 0$. In other words, the dominant kind of phase
 error is having no phase error at all. Thus, the majority vote PEC procedure
 is effective in bringing down the phase error. This is the underlying reason
 why Scheme~A is so powerful that in the limit $N\rightarrow\infty$,
 $e^\textrm{SBMER}\rightarrow 1^-$.

 Fifth, the privacy amplification performed in Scheme~A is based entirely on
 entanglement purification and phase error correction. In fact, the key
 ingredient in reducing the QER used in the proof of
 Theorem~\ref{Thrm:Uncond_Sec_A} is the validity of conditions shown in
 Eq.~(\ref{E:QER_Condition_p2_1}). Nonetheless, there is no need to bring down
 the QER to an exponentially small number. In fact, one may devise an equally
 secure scheme by following the adaptive procedure introduced by Chau in
 Ref.~\cite{sixstateexact}. That is to say, Alice and Bob may switch to a
 concatenated Calderbank-Shor-Steane quantum code when the PEC brings down the
 QER to about 5\%. The strategy of adding an extra step of quantum error
 correction towards the end of the privacy amplification procedure may increase
 the key generation rate. This is because from the proof of
 Theorem~\ref{Thrm:Uncond_Sec_A} together with Eq.~(\ref{E:r_bound_p2}), I
 conclude that in order to bring the QER down to less than $\epsilon$ after $k$
 rounds of LOCC2 EP, Alice and Bob have to choose $r$ and hence the number of
 quantum registers needed in PEC to be $\sim \epsilon c^{2^k}$ for some
 constant $c>1$. In contrast, by randomizing the quantum registers, the QER
 after each application of the Steane's seven quantum register code is reduced
 quadratically whenever the QER is less than about 5\%. Consequently, Alice and
 Bob may increase the key generation rate by performing less rounds of
 LOCC2 EP, choosing $\epsilon \approx 0.01$, and finally adding a few rounds of
 Calderbank-Shor-Steane code quantum error correction procedure.

\section{Reduction To The Prepare-And-Measure Scheme \label{Sec:ShorPre}}
 Finally, I apply the standard Shor and Preskill proof \cite{shorpre} to reduce
 the entanglement-based Scheme~A to a provably secure prepare-and-measure
 scheme in this section. Let me first write down the detail procedures of
 Scheme~B before showing its security.
 
\par\medskip
\begin{enumerate}[\underline{Prepare-and-measure QKD Scheme~B}]
 \item Alice randomly and independently prepares $L\gg 1$ quantum particles in
  the standard basis. She applies one of the following unitary transformation
  to each particle randomly and independently: $I, T, T^2, \ldots , T^N$.
  Alice records the states and transformations she applied and then sends the
  states to Bob. He acknowledges the reception of these particles and then
  applies one of the following transformation to each received particles
  randomly and independently: $I^{-1}, T^{-1}, T^{-2}, \ldots, T^{-N}$. Now,
  Alice and Bob publicly reveal their unitary transformations applied to each
  particle. A particle is kept and is said to be in the set $S_i$ if Alice and
  Bob have applied $T^i$ and $T^{-i}$ to it respectively. Bob measures the
  particles in $S_i$ in the standard basis and records the measurement results.
  \label{PM:Prepare}
 \item Alice and Bob estimate the quantum channel error rate by sacrificing a
  few particles. Specifically, they randomly pick $\mbox{O} ([N+1]^2 \log \{
  [N+1]/\epsilon \} / \delta^2 N^2)$ pairs from each of the $(N+1)$ sets $S_i$
  and publicly reveal the preparation and measured states for each of them. In
  this way, they obtain the estimated channel error rate within standard
  deviation $\delta$ with probability at least $(1-\epsilon)$. If the channel
  error rate is too high, they abort the scheme and start all over again.
  \label{PM:QCTest}
 \item Alice and Bob perform the following privacy amplification procedure.
   \begin{enumerate}
    \item They apply the privacy amplification procedure with two way classical
     communication similar to the ones reported in
     Refs.~\cite{qkd2waylocc,sixstateexact}. Specifically, Alice and Bob
     randomly group their corresponding remaining quantum particles in pairs.
     Suppose the $j$th particle of the $i$th pair was initially prepared in the
     state $|s_{i_j}\rangle$. Then, Alice publicly announces the value $s_{i_1}
     -s_{i_2} \in GF(N)$ for each pair $i$. Similarly, Bob publicly announces
     the value $s'_{i_1}-s'_{i_2}$ where $|s'_{i_j}\rangle$ is the measurement
     result of the $j$th particle in the $i$th pair. They keep one of their
     corresponding registers of the pair only when their announced values the
     corresponding pairs agree. They repeat the above procedure until there is
     an integer $r > 0$ such that a single application of
     step~\ref{Ent:PriAmp}b will bring the quantum channel error rate of the
     resultant particles down to $\epsilon_I /\ell^2$ for a fixed security
     parameter $\epsilon_I > 0$, where $r\ell$ is the number of remaining
     quantum particles they have. They abort the scheme either when $r$ is
     greater than the number of remaining quantum particles they possess or
     when they have used up all their quantum particles in this procedure.
    \item They apply the majority vote phase error correction procedure
     introduced by Gottesman and Lo \cite{qkd2waylocc}. Specifically, Alice and
     Bob randomly divide their corresponding resultant particles into sets
     each containing $r$ particles. They replace each set by the sum of the
     values prepared or measured of the $r$ particles in the set. These
     replaced values are bits of their final secure key string.
   \end{enumerate}
   \label{PM:PriAmp}
\end{enumerate}

\par\medskip
\begin{Thrm}[Based on Shor and Preskill \cite{shorpre}]
 Scheme~A in Section~\ref{Sec:EntScheme} and Scheme~B above are equally secure.
 Thus, conclusions of Theorem~\ref{Thrm:Uncond_Sec_A} is also applicable to
 Scheme~B. \label{Thrm:Uncond_Sec_B} 
\end{Thrm}
\begin{proof} 
 Recall from Ref.~\cite{shorpre} that Alice may measure all her share of
 quantum registers right at step~\ref{Ent:Prepare} in Scheme~A without
 affecting the security of the scheme. Besides, LOCC2 EP and PEC procedures in
 Scheme~A simply permute the measurement basis. More importantly, the final
 secret key generation does not make use of the phase information of the
 transmitted quantum registers. Hence, the Shor-Preskill argument in
 Ref.~\cite{shorpre} can be applied to Scheme~A, giving us an equally secure
 prepare-and-measure Scheme~B above.
\end{proof}

\par\medskip
 From the discussions in Subsection~\ref{Subsec:Error_Rates}, we \emph{should
 not} and \emph{cannot} compare the error tolerant capability of Scheme~B that
 uses unentangled quantum particles of different dimensions as information
 carrier. Nonetheless, we may compare the error tolerant capability of the
 entangled-qubit-based prepare-and-measure QKD scheme derived from Scheme~B
 against the same eavesdropping attack. Recall that in the absence of quantum
 storage, we may regard the transfer of a 16-dimensional quantum particle as
 the transfer of 4 consecutive qubits in the insecure quantum channel.
 Now, I consider the following eavesdropping strategy: Qubits passing through
 the insecure communication channel are partitioned into sets each containing 4
 consecutive qubits. Eve randomly and independently measure each set in the
 standard basis with probability $q$. Suppose $q$ satisfies
\begin{equation}
 0.8292 \approx \frac{3}{10} (5-\sqrt{5}) < q < \frac{68}{1335} (19-\sqrt{5})
 \approx 0.8539 . \label{E:constraint_q}
\end{equation}
 From Lemma~\ref{Lem:UpperBound_QER_SBMER} and Eq.~(\ref{E:BER_QER}), the BER
 caused by this eavesdropping strategy on the entangled-qubit-based
 prepare-and-measure QKD scheme derived from Scheme~B for $N = 2^n$ is given by
 $e^\textrm{BER}_\textrm{Eve}(N) = q(N-1)(N n+2)/[2N n(N+1)]$. Using
 Eqs.~(\ref{E:critical_p2}), (\ref{E:SBMER_QER})--(\ref{E:constraint_q}), I
 conclude that $e^\textrm{BER}_\textrm{Eve}(2) > (5-\sqrt{5})/10$. In other
 words, $e^\textrm{BER}_\textrm{Eve}(2)$ is greater than tolerable BERs of all
 known unentangled-qubit-based prepare-and-measure QKD schemes to date. In
 contrast, $e^\textrm{BER}_\textrm{Eve}(16) < 33(19-\sqrt{5})/1424$. Hence,
 from Theorem~\ref{Thrm:Uncond_Sec_B} together with Eqs.~(\ref{E:critical_p2}),
 (\ref{E:SBMER_QER}) and~(\ref{E:BER_QER}), Scheme~B can generate a provably
 secure key under this eavesdropping attack when $N=16$. Actually, one may
 construct an eavesdropping attack that can be tolerated by the
 entangled-qubit-based prepare-and-measure scheme derived from Scheme~B for a
 fixed $N=2^n \geq 16$ in a similar way. (The strategy is partition the qubits
 into sets each containing $n$ consecutive qubits. Eve makes standard basis
 measurement on each set with probability $q$ chosen from an interval similar
 to the one stated in Eq.~(\ref{E:constraint_q}).) All known
 unentangled-qubit-based prepare-and-measure schemes to date, in contrast,
 cannot generate a provably secure key under the same attack.

 On the other hand, suppose Eve chooses a slightly different strategy by
 measuring randomly and independently a qubit in each set of 4 consecutive
 qubits with probability $q' = 1-[(43+68\sqrt{5})/1335]^{1/4} \approx 0.3817$
 in the standard basis. Under this modified eavesdropping attack, the
 probability that a randomly chosen 4 consecutive qubits are not chosen equals
 $(1-q')^4$ in the limit of large number of qubit transfer. Thus, the BER
 induced by this attack on the entangled-qubit-based prepare-and-measure scheme
 derived from Scheme~B for $N = 16$ is given by $[1-(1-q')^4](N-1)(N n+2)/[2N n
 (N+1)] = 33(19-\sqrt{5})/1424$. This BER rate is just too high for the
 entangled-qubit-based scheme derived from Scheme~B for $N=16$ to handle. In
 contrast, the BER caused by the same eavesdropping attack for the six-state
 scheme equals $q'/3 \approx 0.1272$. This attack, therefore, can be handled
 easily by the unentangled-qubit-based prepared-and-measure QKD scheme
 introduced by Chau in Ref.~\cite{sixstateexact}. To summarize, the
 entangled-qubit-based prepare-and-measure scheme derived from Scheme~B for $N
 > 2$ is more error resilience when dealing with burst type of errors than the
 unentangled-qubit-based prepare-and-measure schemes.

\par\medskip
 Now, I need to point out an important remark on the number of different kinds
 of states Alice have to prepare in Scheme~B. To distribute the key using an
 $N$-level quantum system with $N = 2^n$, Corollary~\ref{Cor:Sequence} tells us
 that $T^k \neq I$ for all $k= 1,2,\ldots ,N$. Therefore, $T^i |j\rangle$'s are
 distinct states for $0\leq i\leq N$ and $j\in GF(N)$. Thus, Scheme~B is a $N
 (N+1)$-state scheme. In contrast, if $N = p^n$ with $p>2$, then $T^{(N+1)/2} =
 -I$ by Corollary~\ref{Cor:Sequence}. Hence, in this case, upon measurement on
 the standard basis, Scheme~B is a $N(N+1)/2$-state scheme. This observation
 suggests that there may be rooms for improving the error tolerance rate of an
 prepare-and-measure QKD scheme involving $N$-dimensional quantum particles for
 an odd $N$.

 Finally, I remark that the lemma below suggests the possibility of a subtle
 relation between Scheme~B and the so-called mutually unbiased bases.

\par\medskip
\begin{Lem}
 If $N = 2^n$, then the bases $\{ |k\rangle \}_{k\in GF(N)}$, $\{ T |k\rangle
 \}_{k\in GF(N)}$, $\{ T^2 |k\rangle \}_{k\in GF(N)}, \ldots ,$ $\{ T^N |k
 \rangle \}_{k\in GF(N)}$ are mutually unbiased. While if $N = p^n$ with $p>2$,
 the bases $\{ |k\rangle \}_{k\in GF(N)}$, $\{ T |k\rangle \}_{k\in GF(N)},
 \ldots ,$ $\{ T^{(N+1)/2} |k\rangle \}_{k\in GF(N)}$ are mutually unbiased.
 \label{Lem:MUB}
\end{Lem}
\begin{proof}
 I shall only consider the case when $N = 2^n$. The other case can be proven in
 the same way. Let $0 \leq i < i' \leq N$. I consider the equation
\begin{equation}
 \langle k'| {T^i}^\dag T^{i'} |k\rangle = \langle 0| Z_j X_{-k'} T^{i'-i} X_k
 |0\rangle , \label{E:MUB1}
\end{equation}
 which holds for all $j\in GF(N)$. Since $0 < i' - i \leq N$,
 Corollary~\ref{Cor:Sequence} implies that $M(T^{i' - i})$ is in the form
 $\left[ \begin{array}{cc} a & b \\ b & c \end{array} \right]$ for some $b\neq
 0$. Therefore, applying Eqs.~(\ref{E:Algebra_X_Z}) and~(\ref{E:XZT}) to the
 right hand side of Eq.~(\ref{E:MUB1}) gives an expression proportional to
 $\langle 0| T^{i' - i} X_{k-k'a+j b} Z_{-k'b + j c} |0\rangle = \langle 0 |
 T^{i' - i} |k-k'a+j b\rangle$. More importantly, the magnitude of the
 proportionality constant equals 1 for all $j,k,k'\in GF(N)$. Hence, $| \langle
 k'| T^i |k\rangle |^2 = | \langle k'' | T^i |k\rangle |^2$ for all $k,k',k''
 \in GF(N)$ whenever $0 < i \leq N$. Hence, $\{ |k\rangle \}_{k\in GF(N)}$, $\{
 T |k \rangle \}_{k\in GF(N)}, \ldots,$ $\{ T^N |k\rangle \}_{k\in GF(N)}$ are
 mutually unbiased.
\end{proof}

\par\medskip\indent
 Since the maximum number of mutually unbiased bases equals $(N+1)$ for any
 prime power $N$ \cite{mub1,mub2,mub3}, the construction in Scheme~B provides a
 simple way to build such mutually unbiased bases for $N = 2^n$. Perhaps one
 may build a more error tolerant QKD scheme using mutually unbiased bases for
 the case of an odd prime power $N$.

\section{Discussions \label{Sec:Dis}}
 In summary, I have introduced a prepared-and-measured QKD scheme (Scheme~B)
 and proved its unconditional security. In particular, I show that for a
 sufficiently large Hilbert space dimension of quantum particles $N$ used,
 Scheme~B generates a provably secure key close to 100\% SBMER or 50\% BER.
 This result demonstrates the advantage of using unentangled higher dimensional
 quantum particles as signal carriers in QKD.

 A variation to the theme is worth discussing. Suppose Alice can only send
 qubits. Besides, she can entangle the qubits but she cannot store them. Then,
 she may group $n$ qubits together as a $2^n$-dimensional system and apply
 Scheme~B. Under this situation, Scheme~B can generate a provably secure key
 under certain eavesdropping attack whenever $n\geq 4$. In contrast, no
 unentangled-qubit-based prepare-and-measure QKD scheme known to date can
 tolerate the same eavesdropping attack. Nonetheless, there exists another
 eavesdropping attack that Scheme~B cannot tolerate unless $N=2$. Recall that
 Scheme~B is equivalent to the unentangled-qubit-based prepare-and-measure
 scheme proposed by Chau in Ref.~\cite{sixstateexact}. Therefore, the ability
 to create, transfer but not to store entangle qubits is advantageous in
 quantum cryptography using certain quantum channels with burst errors.

 There is a tradeoff between the error tolerance rate and key generation
 efficiency, however. It is clear from the proof of
 Theorem~\ref{Thrm:Uncond_Sec_A} that $r$ and hence the number of quantum
 particle transfer from Alice and Bob $L$ scales as $2^k$. Besides, the
 probability that the measurement results agree and hence the control quantum
 register pairs are kept in LOCC2 EP equals $\approx 1/N$ in the worst case.
 As a result, while the Scheme~B is highly error-tolerant, it generates a
 secret key with exponentially small efficiency in the worst case scenario.
 Fortunately, the adaptive nature of Scheme~B makes sure that this scenario
 will not happen when the error rate of the channel is small. To conclude, in
 most practical situations, Alice and Bob should choose the smallest possible
 $N$ whose corresponding $e^\textrm{SBMER}$ is slightly larger than the channel
 standard basis measurement error rate. In this way, they can almost surely
 generate their provably secure key at the highest possible rate.

 As I have noted in Section~\ref{Sec:ShorPre}, there may be room for improving
 the error tolerance rate in the case $p>2$ since Scheme~B uses only $N(N+1)/2$
 different quantum states in signal transmission. It is instructive to explore
 such a possibility.

\section*{Acknowledgments}
 This work is supported in part by the Outstanding Young Researcher Award of
 the University of Hong Kong. The author would like to thank H.-K. Lo for
 sharing with him his preprint with D. Gottesman \cite{qkd2waylocc} prior to
 its public dissemination. A critical reading of an earlier draft by
 Debbie~Leung is also gratefully acknowledged.
\bibliographystyle{ieeetr}
\bibliography{qc30.3}

\begin{thebibliography}{10}

\bibitem{NC}
M.~A. Nielsen and I.~L. Chuang, {\em Quantum Computation And Quantum
  Information}.
\newblock Cambridge: CUP, 2000.
\newblock p.~586.

\bibitem{biasedbb84}
H.-K. Lo, H.~F. Chau, and M.~Ardehali, ``Efficient quantum key distribution
  scheme and proof of its unconditional security,'' 2001.
\newblock (quant-ph/0011056v2), to appear in J.\ Crypt.

\bibitem{lochauqkdsec}
H.-K. Lo and H.~F. Chau, ``Unconditional security of quantum key distribution
  over arbitrarily long distances,'' {\em Science}, vol.~283, pp.~2050--2056,
  1999.
\newblock As well as the supplementary material available at
  {\texttt{http://www.sciencemag.org/feature/data/984035.shl}}.

\bibitem{mayersjacm}
D.~Mayers, ``Unconditional security in quantum cryptography,'' {\em J.\ Assoc.\
  Comp.\ Mach.}, vol.~48, pp.~351--406, 2001.
\newblock See also his preliminary version in D. Mayers, \textit{Advances in
  Cryptology --- Proceedings of Crypto'96} (Springer Verlag, Berlin, 1996),
  pp.~343--357.

\bibitem{gottloreview}
D.~Gottesman and H.-K. Lo, ``From quantum cheating to quantum security,'' {\em
  Phys.\ Today}, vol.~53, no.~11, pp.~22--27, 2000.
\newblock And references cited therein.

\bibitem{gisinreview}
N.~Gisin, G.~Ribordy, W.~Tittel, and H.~Zbinden, ``Quantum cryptography,'' {\em
  Rev.\ Mod.\ Phys.}, vol.~74, pp.~145--195, 2002.
\newblock And references cited therein.

\bibitem{bb84}
C.~H. Bennett and G.~Brassard, ``Quantum cryptography: Public key distribution
  and coin tossing,'' in {\em Proceedings of the IEEE International Conference
  on Computers, Systems and Signal Processing}, (New York), pp.~175--179,
  Bangalore, India, IEEE, 1984.

\bibitem{sixstate}
D.~Bru{\ss}, ``Optimal eavesdropping in quantum cryptography with six states,''
  {\em Phys.\ Rev.\ Lett.}, vol.~81, pp.~3018--3021, 1998.

\bibitem{cont1}
T.~C. Ralph, ``Continuous variable quantum cryptography,'' {\em Phys.\ Rev.\
  A}, vol.~61, pp.~010303(R):1--4, 2000.

\bibitem{cont2}
M.~Hillery, ``Quantum cryptography with sequeezed states,'' {\em Phys.\ Rev.\
  A}, vol.~61, pp.~022309:1--8, 2000.

\bibitem{squeez}
D.~Gottesman and J.~Preskill, ``Secure quantum key distribution using squeezed
  states,'' {\em Phys.\ Rev.\ A}, vol.~63, pp.~022309:1--18, 2001.

\bibitem{threestate}
H.~Bechmann-Pasquinucci and A.~Peres, ``Quantum cryptography with 3-state
  systems,'' {\em Phys.\ Rev.\ Lett.}, vol.~85, pp.~3313--3316, 2000.

\bibitem{alph}
H.~Bechmann-Pasquinucci and W.~Tittel, ``Quantum cryptography using larger
  alphabets,'' {\em Phys.\ Rev.\ A}, vol.~61, pp.~062308:1--6, 2000.

\bibitem{highdim}
M.~Bourennane, A.~Karlsson, and G.~Bj{\"o}rk, ``Quantum key distribution using
  multilevel encoding,'' {\em Phys.\ Rev.\ A}, vol.~64, pp.~012306:1--5, 2001.

\bibitem{dlevel}
N.~J. Cerf, M.~Bourennane, A.~Karlsson, and N.~Gisin, ``Security of quantum key
  distribution using d-level systems,'' {\em Phys.\ Rev.\ Lett.}, vol.~88,
  pp.~127902:1--4, 2002.

\bibitem{highdimmore}
M.~Bourennane, A.~Karlsson, G.~Bj{\"o}rk, N.~Gisin, and N.~J. Cerf, ``Quantum
  key distribution using multilevel encoding: security analysis,'' {\em J.\
  Phys.:\ A}, vol.~35, pp.~10065--10076, 2002.

\bibitem{evethreestate}
D.~Bru{\ss} and C.~Macchiavello, ``Optimal eavesdropping in cryptography with
  three-dimensional quantum states,'' {\em Phys.\ Rev.\ Lett.}, vol.~88,
  pp.~127901:1--4, 2002.

\bibitem{biham}
E.~Biham, M.~Boyer, P.~O. Boykin, T.~Mor, and V.~Roychowdhury, ``A proof of the
  security of quantum key distribution,'' in {\em Proceedings of the 32nd
  Annual ACM Symposium on Theory of Computing (STOC2000)}, (New York),
  pp.~715--724, ACM Press, 2000.

\bibitem{bdsw}
C.~H. Bennett, D.~A. DiVincenzo, J.~A. Smolin, and W.~K. Wootters,
  ``Mixed-state entanglement and quantum error correction,'' {\em Phys.\ Rev.\
  A}, vol.~54, pp.~3824--3851, 1996.

\bibitem{shorpre}
P.~W. Shor and J.~Preskill, ``Simple proof of security of the \uppercase{BB84}
  quantum key distribution protocol,'' {\em Phys.\ Rev.\ Lett.}, vol.~85,
  pp.~441--444, 2000.

\bibitem{sixstateproof}
H.-K. Lo, ``Proof of unconditional security of six-state quantum key
  distribution scheme,'' {\em Quant.\ Inform.\ and\ Comp.}, vol.~1, no.~2,
  pp.~81--94, 2001.

\bibitem{qkd2waylocc}
D.~Gottesman and H.-K. Lo, ``Proof of security of quantum key distribution with
  two-way classical communications,'' {\em IEEE\ Trans.\ Inf.\ Theo.}, vol.~49,
  pp.~457--475, 2003.

\bibitem{sixstateexact}
H.~F. Chau, ``Practical scheme to share a secret key through a quantum channel
  with a 27.5\% bit error rate,'' {\em Phys.\ Rev.\ A}, vol.~66,
  pp.~060302(R):1--4, 2002.

\bibitem{QECCHammingBound}
D.~Gottesman, ``Class of quantum error-correcting codes saturating the quantum
  \uppercase{H}amming bound,'' {\em Phys.\ Rev.\ A}, vol.~54, pp.~1862--1868,
  1996.

\bibitem{knill}
A.~Ashikhmin and E.~Knill, ``Non-binary quantum stabilizer codes,'' {\em IEEE\
  Trans.\ Inf.\ Theo.}, vol.~47, pp.~3065--3072, 2001.

\bibitem{finitefieldbook}
R.~Lidl and H.~Neiderreiter, {\em Introduction to finite fields and their
  applications}.
\newblock Melbourne: CUP, revised~ed., 1994.

\bibitem{genpur}
G.~Alber, A.~Delgado, N.~Gisin, and I.~Jex, ``Efficient bipartite quantum state
  purification in arbitrary dimensional \uppercase{H}ilbert spaces,'' {\em J.\
  Phys.:A}, vol.~34, pp.~8821--8833, 2001.

\bibitem{multidist}
N.~L. Johnson, S.~Kotz, and N.~Balakrishnan, {\em Discrete Multivariate
  Distributions}.
\newblock New York: Wiley, 1997.
\newblock chap.~39.

\bibitem{hd_ftqc}
D.~Gottesman, ``Fault-tolerant quantum computation with higher-dimensional
  systems,'' {\em Chaos,\ Solitons\ \&\ Fractals}, vol.~10, pp.~1749--1758,
  1999.

\bibitem{Roman}
S.~Roman, {\em Coding And Information Theory}.
\newblock Berlin: Springer, 1992.
\newblock p.~26.

\bibitem{mub1}
W.~K. Wootters and B.~D. Fields, ``Optimal state-determination by mutually
  unbiased measurements,'' {\em Ann.\ Phys.}, vol.~191, pp.~363--381, 1989.

\bibitem{mub2}
J.~Lawrence, C.~Brukner, and A.~Zeilinger, ``Mutually unbiased binary
  observable sets on $\uppercase{N}$ qubits,'' {\em Phys.\ Rev.\ A}, vol.~65,
  pp.~032320:1--5, 2002.

\bibitem{mub3}
S.~Bandyopadhyay, P.~O. Boykin, V.~Roychowdhury, and F.~Vatan, ``A new proof
  for the existence of mutually unbiased bases,'' {\em Algorithmica}, vol.~34,
  pp.~512--528, 2002.

\end{thebibliography}
\end{document}